\pdfoutput=1
\documentclass[aps, twocolumn, floatfix, superscriptaddress]{revtex4-1}
\usepackage{graphicx}
\DeclareGraphicsExtensions{.pdf}
\usepackage{amsmath,amssymb,bbold,bm,color}
\usepackage{float}
\usepackage{epstopdf}
\usepackage{tabularx}
\usepackage{braket}
\usepackage{booktabs}
\usepackage{array}
\usepackage{blindtext}
\usepackage[colorlinks=true,breaklinks=true,linkcolor=blue,anchorcolor=red,citecolor=blue,urlcolor=blue]{hyperref}

\DeclareMathAlphabet\mathbfcal{OMS}{cmsy}{b}{n}
\usepackage{hhline}
\newcolumntype{L}[1]{>{\raggedright\let\newline\\\arraybackslash\hspace{0pt}}m{#1}}
\newcolumntype{C}[1]{>{\centering\let\newline\\\arraybackslash\hspace{0pt}}m{#1}}
\newcolumntype{R}[1]{>{\raggedleft\let\newline\\\arraybackslash\hspace{0pt}}m{#1}}

\begin{document}

\title{Pseudo Landau levels, negative strain resistivity, and enhanced thermopower in twisted graphene nanoribbons}

\author{Zheng Shi}
\affiliation{Dahlem Center for Complex Quantum Systems and Physics Department, Freie Universit\"at Berlin, 14195 Berlin, Germany}

\affiliation{Institute for Quantum Computing and Department of Physics and Astronomy, University of Waterloo, Waterloo, Ontario, N2L 3G1, Canada}

\author{Hai-Zhou Lu}
\affiliation{Institute for Quantum Science and Engineering and Department of Physics, Southern University of Science and Technology, Shenzhen 518055, China}
\affiliation{Shenzhen Key Laboratory of Quantum Science and Engineering, Shenzhen 518055, China}

\author{Tianyu Liu}
\email{tliu@pks.mpg.de}
\affiliation{Max-Planck-Institut f\"ur Physik komplexer Systeme, 01187 Dresden, Germany}

\begin{abstract} 
As a canonical response to the applied magnetic field, the electronic states of a metal are fundamentally reorganized into Landau levels. In Dirac metals, Landau levels can be expected without magnetic fields, provided that an inhomogeneous strain is applied to spatially modulate electron hoppings in a way similar to the Aharonov-Bohm phase. We here predict that a twisted zigzag nanoribbon of graphene exhibits strain-induced pseudo Landau levels of unexplored but analytically solvable dispersions at low energies. The presence of such dispersive pseudo Landau levels results in a negative strain resistivity characterizing the $(1+1)$-dimensional chiral anomaly if partially filled and can greatly enhance the thermopower when fully filled.
\end{abstract}

\date{\today}
\maketitle

\section{Introduction}
\label{s1}
A magnetic field applied to a metal can quantize the orbital motion of electrons and populate them on discrete energy bands known as the Landau levels (LLs)  \cite{landau1930}, which are responsible for a number of transport properties. When the applied magnetic field is scanned, LLs can successively pass through the Fermi surface, giving rise to quantum oscillations such as the Shubnikov-de Haas effect \cite{SdH1, SdH2} and the de Haas-van Alphen effect \cite{dAvH}. In recently discovered topological semimetals \cite{armitage2018, wan2011, burkov2011, young2012, wang2012, wang2013}, the presence of LLs accounts for the non-conservation of chiral charge transport, i.e., the chiral anomaly \cite{adler1969, bell1969, nielsen1983}, which is observable through a negative longitudinal magnetoresistivity \cite{huang2015, kim2013, xiong2015, zhang2016, son2013, burkov2015, li2016} resulting from the chiral magnetic effect \cite{son2013, burkov2015, li2016, fukushima2008}. The integer quantum Hall effect in the massive two-dimensional electron gas \cite{klitzing1980, laughlin1981, klitzing1986}, the half-integer quantum Hall effect in graphene \cite{gusynin2005, gusynin2006, peres2007, krstajic2011, zhang2005, novoselov2005}, and the fractional quantum Hall effect in incompressible quantum liquids \cite{tsui1982, laughlin1983, jain1989, jain1990, stormer1999} all derive from  particular filling factors of LLs.

Landau levels, remarkably, have been proposed to exist in elastically strained Dirac metals in the absence of magnetic fields \cite{vozmediano2010, ilan2020, arjona2017, castro2017}, leading to strain-induced transport phenomena such as quantum oscillations \cite{liu2017a, liu2020a}, quantum anomalies \cite{pikulin2016, grushin2016}, and Hall-like effects \cite{guinea2010a, roy2013, venderbos2016, sela2020}, similar to those in the context of the regular magnetotransport. Such strain-induced pseudo-Landau levels (pLLs) have been experimentally observed by scanning tunneling spectroscopy (STS) in nanobubbles \cite{levy2010, lu2012} and nanoripples \cite{yeh2011, li2015} of graphene and directly imaged by angle-resolved photoemission spectroscopy (ARPES) in wafer-scale epitaxially grown graphene on a silicon carbide (SiC) substrate with nanoprisms \cite{nigge2019}. To interpret the transport experiments involving pLLs, one should ideally understand the pLL dispersions. However, as presumably the most flexible Dirac metal and thus the most promising experimental venue, graphene requires great experimental effort in fine-tuning the strain in a triaxial pattern for the induction of the regular flat Dirac-Landau levels \cite{guinea2010a, settnes2016, low2010}, while simple strain patterns such as those arising from bending \cite{guinea2010b, chang2012, stuij2015} or twisting \cite{zhang2014} a graphene nanoribbon (GNR) produce complicated dispersive pLLs.

In this paper, we propose a general method based on the band theory to analytically derive the band structure of pLLs induced in a twisted zigzag GNR and then use the resolved pLL dispersions to interpret the transport signatures of the twisted GNR. In Sec.~\ref{s2}, we show that the twisted GNR exhibits a bulk zero mode which is the strain-induced zeroth pLL (pLL$_0$) by nature. By linearizing the Hamiltonian of the twisted GNR in the vicinity of the bulk zero mode, i.e., the pLL guiding center, we derive the dispersions of the pLLs at low energies. In Sec.~\ref{s3}, we study the low-energy transport of the twisted GNR in the framework of the semiclassical Boltzmann formalism and elucidate that the dispersive pLLs engender a negative strain resistivity if partially filled and can enhance the thermopower if fully filled. Section~\ref{s4} concludes the paper and envisages a few other venues to which our general method may be applied.

\section{Electronic structure of the twist-induced pseudo-Landau levels}
\label{s2}
\begin{figure}[tb]
\includegraphics[width = 8.6cm]{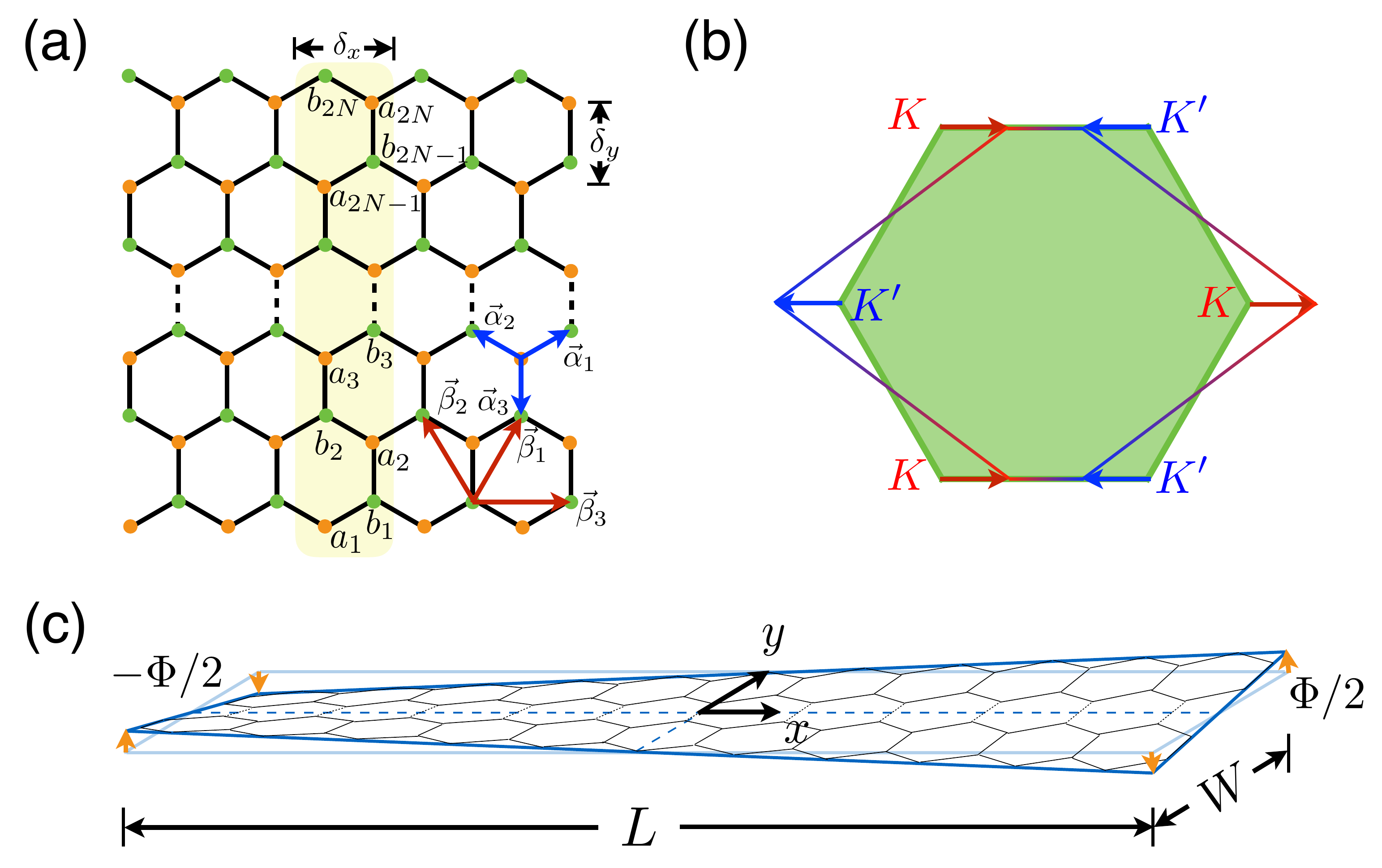}
\caption{(a) Graphene nanoribbon (GNR) with zigzag edges along the $x$ direction. Denoting the honeycomb lattice constant by $a$, we have $\delta_x=\sqrt 3a$ and $\delta_y=\tfrac{3}{2}a$. The nearest neighbor vectors (blue) are thus $(\bm \alpha_1, \bm \alpha_2, \bm \alpha_3) = (\tfrac{1}{2}\delta_x \hat x + \tfrac{1}{3}\delta_y\hat y, -\tfrac{1}{2}\delta_x \hat x + \tfrac{1}{3}\delta_y\hat y, -\tfrac{2}{3}\delta_y \hat y)$, and the next nearest neighbor vectors (red) are $(\bm \beta_1, \bm \beta_2, \bm \beta_3) = (\tfrac{1}{2}\delta_x\hat x + \delta_y\hat y, -\tfrac{1}{2}\delta_x\hat x + \delta_y\hat y, \delta_x\hat x)$. In the presence of open boundary conditions along the $y$ direction, the unit cell of the zigzag GNR is chosen as the shaded area (yellow) containing $2N$ sites on each sublattice. (b) The hexagonal Brillouin zone (green) of the honeycomb lattice in panel (a) contains two Dirac cones at the BZ corners labeled by $K$ and $K'$, respectively. In the presence of spatially uniform anisotropic hoppings $(t_1, t_2, t_3) = (t+\delta t, t+\delta t, t)$, where $\delta t<0$, the two Dirac cones are oppositely displaced away from BZ corners as illustrated by the red and blue arrows. (c) A twisted GNR (blue) is obtained by applying a torsional strain that rotates the right (left) edge of the undeformed GNR (light blue) of length $L$ and width $W$ by a small angle of $\Phi/2$ ($-\Phi/2$).  } \label{fig1}
\end{figure}

We start from the strain-free tight-binding Hamiltonian of graphene with only nearest neighbor hopping terms
\begin{equation} \label{H_inf}
H_0=\sum_{\bm r} \sum_i t_i a_{\bm r}^\dagger b_{\bm r + \bm \alpha_i} +\text{H.c.},
\end{equation}
where $a_{\bm r} (b_{\bm r})$ is the electronic annihilation operator at site $\bm r=(x,y)$ belonging to the $A(B)$ sublattice of the honeycomb lattice with lattice constant $a=0.142$ nm \cite{neto2009, dassarma2011, goerbig2011}, and $\bm \alpha_i$ is the $i$th nearest neighbor vector along which the hopping is $t_i$ independent of $\bm r$. In the following, unless otherwise specified, we choose the $x$ direction to be parallel to the zigzag edges [Fig.~\ref{fig1}(a)], so that $(\bm \alpha_1, \bm \alpha_2, \bm \alpha_3) = (\tfrac{\sqrt 3}{2} a \hat x + \tfrac{1}{2} a \hat y, -\tfrac{\sqrt 3}{2}a \hat x + \tfrac{1}{2} a \hat y, -a \hat y)$. This tight-binding Hamiltonian encodes two energy bands $\varepsilon(\bm k)=\pm |\sum_i t_i e^{i\bm k \cdot \bm \alpha_i}|$, which exhibit Dirac cones at the Brillouin zone (BZ) corners \cite{rostami2012} for isotropic hoppings $t_{i=1,2,3}=t$. For anisotropic hoppings $(t_1, t_2, t_3) = (t+\delta t, t+\delta t, t)$, the Dirac cones are translated from the BZ corners to $\bm k_W^\pm=[\pm\tfrac{2}{\sqrt 3 a} \arccos(-\tfrac{1}{2}\tfrac{t}{t+\delta t}), 0]$, as illustrated in Fig.~\ref{fig1}(b).

An elastic strain can deform the crystalline lattice of graphene, break the translational symmetry, and spatially modulate the hoppings through the empirical formula \cite{pereira2009} 
\begin{equation} \label{t_sub}
t_i \rightarrow t_i \exp \{g [1-\tilde \alpha_i(\bm r)/\alpha_i] \},
\end{equation}
where $g=3.37$ is the Gr\"uneisen parameter \cite{pereira2009, carrillo2014}, $\tilde \alpha_i(\bm r)$ is the strain-modulated spacing between a chosen lattice site at $\bm r$ and its $i$th nearest neighbor, and $\alpha_i=a$ is the strain-free counterpart of $\tilde \alpha_i(\bm r)$ as illustrated in Fig.~\ref{fig1}(a). For the simple twist lattice deformation [Fig.~\ref{fig1}(c)] characterized by the parameter $\lambda = \Phi/L$ that measures the rotational angle of the GNR unit cell [Fig.~\ref{fig1}(a)] per unit length along the $x$ direction, we have $\tilde \alpha_i(\bm r) \approx (\alpha_i^2+\lambda^2 \alpha_{i,x}^2 y^2)^{1/2}$ for a sufficiently small twist $\lambda a \ll 1$ \cite{liu2020b}. We note that $\tilde \alpha_i(\bm r)$ preserves the modified $x$-direction translational symmetry $\Pi(\delta_x) = T(\delta_x) R(\lambda \delta_x)$, which should be defined as a regular translation by $\delta_x$ along the $x$ direction combined with a counter-clockwise rotation by an angle of $\lambda \delta_x$ around the $x$ axis [Fig.~\ref{fig1}(c)]. Applying Fourier transform in the $x$ direction, we find that the modulation [Eq.~(\ref{t_sub})] yields for the twisted zigzag GNR a tight-binding Hamiltonian
\begin{equation} \label{H_ribbon}
H=\sum_{k_x, y} a_{k_x,y-\delta_y/3}^\dagger \big[ 2t(y) \cos(\tfrac{1}{2} k_x\delta_x ) + t \hat s_{-\delta_y} \big]  b_{k_x, y} + \text{H.c.},
\end{equation}
where $\hat s_{-\delta_y} b_{k_x,y} =  b_{k_x,y-\delta_y}$ and $t(y) = t \exp \{ g [1-(1+\tfrac{3}{4}\lambda^2 y^2)^{1/2}] \}$ corresponding to the hoppings along $\bm \alpha_1$ and $\bm \alpha_2$, while the hopping along $\bm \alpha_3$ is preserved as $t$. Therefore, the effect of the twist is similar to that of the aforementioned anisotropy ($t_1=t_2 \neq t_3$), relocating the Dirac cones but in a space-dependent fashion. 

To scrutinize this relocation, we take the continuum limit such that the shift operator can be estimated through linearization as $\hat s_{\pm \delta_y} \approx 1 \pm \delta_y \tfrac{d}{dy}$, which leads to the Bloch Hamiltonian for the twisted GNR
\begin{equation} \label{H_Bloch}
\mathcal{H}_{k_x, y} = \big[2 t(y) \cos (\tfrac{1}{2}k_x \delta_x )+ t\big] \tau^x -i t \delta_y \frac{d}{dy} \tau^y.
\end{equation}
The nanoribbon tight-binding Hamiltonian [Eq.~(\ref{H_ribbon})] then becomes $H=\sum_{k_x,y} \psi_{k_x,y}^\dagger \mathcal{H}_{k_x, y} \psi_{k_x,y}$ with Pauli matrices $\tau^x$ and $\tau^y$ acting on $\psi_{k_x,y} = (a_{k_x, y}, b_{k_x, y+\delta_y/3 })^T$. If $t(\tfrac{W}{2}) < \tfrac{1}{2} t$, for any given momentum $\tfrac{4\pi}{3\delta_x} \leq k_x \leq \tfrac{8\pi}{3\delta_x}$, we can always find within the twisted GNR a pair of spatial coordinates 
\begin{equation} \label{y0}
y_0= \pm \frac{2}{\sqrt 3 \lambda} \sqrt {\big\{1+\tfrac{1}{g} \ln \big[ -2 \cos ( \tfrac{1}{2} k_x\delta_x) \big] \big\}^2 -1 },
\end{equation}
at which the first term in Eq.~(\ref{H_Bloch}) changes sign. Therefore, for each choice of sign in Eq.~(\ref{y0}), there exists a bulk zero mode $\Psi_0(y)$, which is an even function of $y-y_0$, satisfying $\mathcal H_{k_x, y} \Psi_0(y)|_{y=y_0}=0$. Such a bulk zero mode comprises the twist-displaced Dirac points associated with different values of $y_0$. 

To better understand the nature of these bulk zero modes, we investigate the spectrum of the Bloch Hamiltonian [Eq.~(\ref{H_Bloch})] at low energies. Because of the exponentially decaying $t(y)$, analytically resolving the eigenvalues of $\mathcal{H}_{k_x, y}$ is usually not feasible. However, if the twist is sufficiently small, $t(y)$ varies slowly on the lattice scale and can be well estimated in the vicinity of the bulk zero mode through the linearization $t(y) \approx t(y_0)[1-\Omega_{y_0} (y-y_0)/t]$, where $\Omega_{y_0}=\tfrac{3}{4} \lambda^2 y_0 g t/ [1+\tfrac{3}{4}\lambda^2 y_0^2]^{1/2}$. This helps simplify the Bloch Hamiltonian [Eq.~(\ref{H_Bloch})] into a minimally coupled Dirac Hamiltonian
\begin{equation} \label{H_Dirac}
\mathcal H_{k_x,y} \approx \Omega_{y_0} (y-y_0) \tau^x - it\delta_y \frac{d}{dy} \tau^y,
\end{equation}
whose eigenvalues are dispersive pLLs (see Appendix~\ref{a1} for detailed derivations)
\begin{equation} \label{pLL}
\begin{split}
&\epsilon_n^\pm(k_x) = \pm \text{sgn}(t\delta_y \Omega_{y_0}) \sqrt{|2n t\delta_y \Omega_{y_0} |} 
\\
&= \pm \frac{3t}{2} \sqrt{n g \lambda a \frac{\tfrac{2}{\sqrt 3} \sqrt{ \big\{1+\tfrac{1}{g} \ln [2\cos(\tfrac{1}{2}k_x\delta_x)] \big\}^2-1}}{1+\tfrac{1}{g} \ln [2\cos(\tfrac{1}{2}k_x\delta_x)]}}, 
\end{split}
\end{equation}
where the momentum dependence is incorporated through $\Omega_{y_0}$ by making use of Eq.~(\ref{y0}) and mapping its domain of definition from $[\tfrac{4\pi}{3\delta_x}, \tfrac{8\pi}{3\delta_x}]$ to $[-\tfrac{2\pi}{3\delta_x}, \tfrac{2\pi}{3\delta_x}]$, which is located in the first BZ of the twisted GNR. We mention that these pLLs are bounded from above: By noting that the pLL energies cannot exceed the merging points of the two Dirac cones, i.e. the Lifshitz transition points $\varepsilon_{\text{Lif}}^\pm=\pm t$, one can estimate from Eq.~(\ref{pLL}) the maximal pLL index as $n_{\text{max}}= \lfloor \tfrac{2 \sqrt 3}{9g \lambda a}[1-\tfrac{1}{(1+g^{-1}\ln2)^2}]^{-1/2} \rfloor$, where $\lfloor \cdot \rfloor$ is the floor function; but we emphasize that Eq.~(\ref{pLL}) fails before $n$ reaches $n_{\text{max}}$ because it is based on the linearization of the Bloch Hamiltonian [Eq.~(\ref{H_Bloch})]. It is worth noting that the pLLs characterized by Eq.~(\ref{pLL}) are doubly degenerate due to the multivaluedness of $\text{sgn}(t\delta_y \Omega_{y_0})$ associated with the contributions from both the upper ($y_0>0$) and the lower ($y_0<0$) sectors of the GNR. The bulk zero modes of the nanoribbon tight-binding Hamiltonian [Eq.~(\ref{H_ribbon})] connecting Dirac points $K$ and $K'$ are none other than the doubly degenerate pLL$_0$ and belong to the family of pLLs.

\begin{figure} [tb]
\includegraphics[width = 8.6cm]{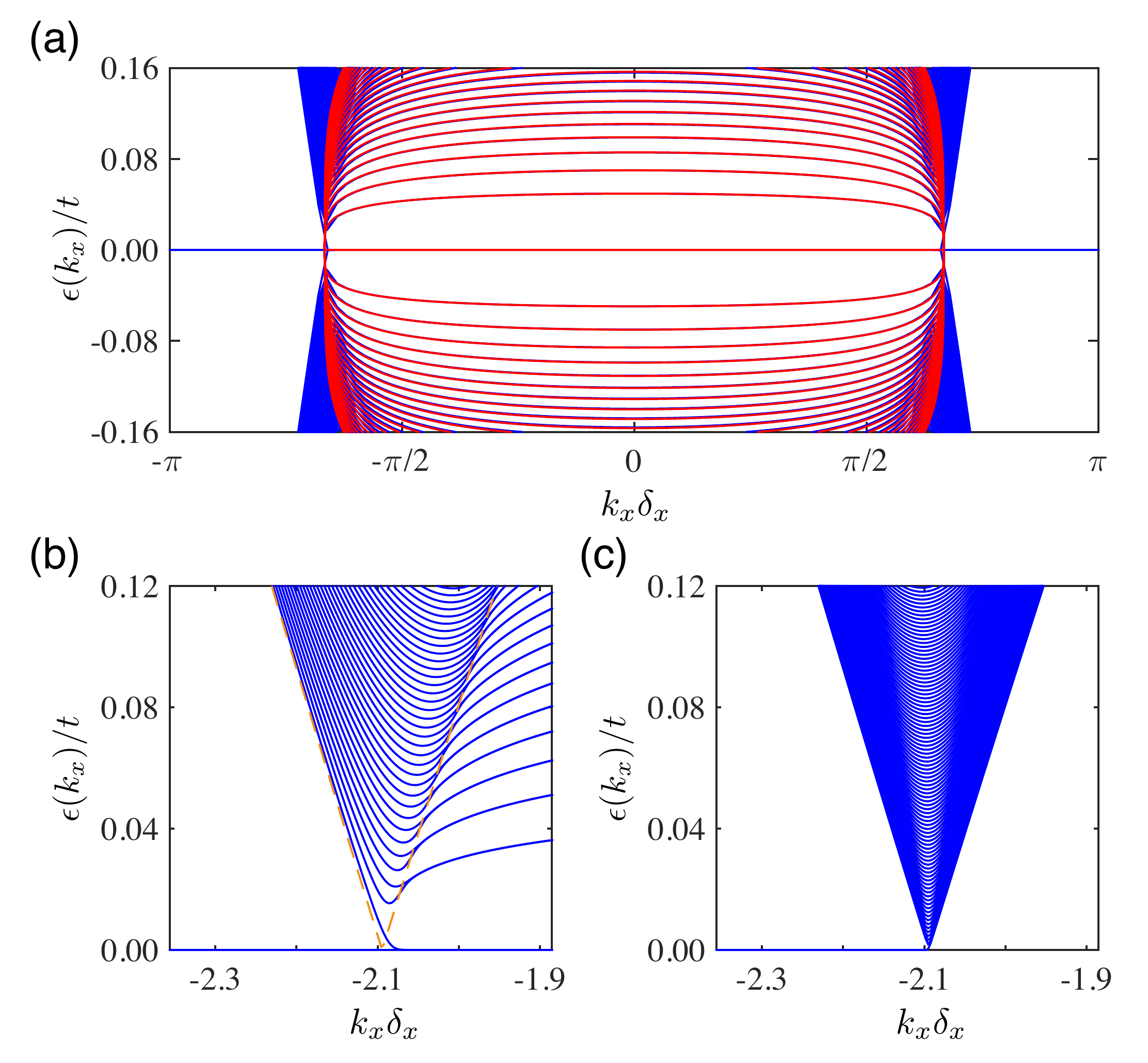}
\caption{Band structure of the twisted GNR with $N=1200$. (a) The low-energy spectrum (blue) is composed of dispersive pLLs and two Dirac cones from which these pLLs emerge. The dispersions of these pLLs are accurately captured by Eq.~(\ref{pLL}) (red), which is overlaid on the energy bands. The twist parameter adopted is $\lambda=0.0005a^{-1}$, at which the maximal C-C bond elongation appearing at the edges of the GNR is $27\%$. Such strain, though large, should be sustainable in graphene \cite{warner2012}. (b) A closer look of the left Dirac cone in panel (a). The dashed curve is the envelope of the cone $\varepsilon_D^\pm (k_x)= \pm [2\cos(\tfrac{1}{2}k_x\delta_x)-1]t$. Each pLL is doubly degenerate, formed by the confluence of two bulk bands at the right sector of the envelope. (c) Dirac cone located at $K$ of an undeformed GNR of the same width harbors many more bands than that in panel (b) in the same energy window. } \label{fig2}
\end{figure}

We have applied numerical simulations to the nanoribbon tight-binding Hamiltonian [Eq.~(\ref{H_ribbon})] and find that the dispersive pLLs are indeed well captured by Eq.~(\ref{pLL}) at low energies [Fig.~\ref{fig2}(a)]. We also notice that the pLLs are actually bounded between two Dirac cones in the momentum dimension. This is because when $k_x$ approaches the Dirac points, $y_0$ eventually becomes comparable to the spatial extent of the pLL wave functions; the wave functions then begin to overlap, lose their degeneracy and recombine into dispersive snake states which constitute the Dirac cones \cite{liuyang2015} [Fig.~\ref{fig2}(b)]. While the energy bands inside the Dirac cones are qualitatively analogous to those in strain-free Dirac cones [Fig.~\ref{fig2}(c)], the strained Dirac cones host far fewer bands, because the strain confines the snake states to the center of the GNR, thereby reducing the effective width of the system.

Besides the pLLs and the bulk states in the Dirac cones of the twisted GNR, we also find a pair of flat bands traversing the whole BZ in Fig.~\ref{fig2}(a). We argue that such bands can only be the edge states of the GNR. To justify this argument, we analyze the nanoribbon tight-binding Hamiltonian [Eq.~(\ref{H_ribbon})] from another point of view by treating it as a Su-Schrieffer-Heeger (SSH) model \cite{su1979} with bipartite hoppings $2t(y)\cos(\tfrac{1}{2}k_x\delta_x)$ and $t$. Between the Dirac points, the GNR unit cell is divided into three segments by the two domain walls at $y_0$ such that the outer two segments (i.e., $|y|>|y_0|$) are topological SSH chains with $2t(y)\cos(\tfrac{1}{2}k_x\delta_x)<t$, while the inner segment (i.e., $|y|<|y_0|$) is a trivial SSH chain with $2t(y)\cos(\tfrac{1}{2}k_x\delta_x)>t$. Therefore, there are in total four ``end states'' associated with the outer two segments. Two of them are the doubly degenerate pLL$_0$ centered at $y_0$, while the other two located at $\pm \tfrac{W}{2}$ constitute the edge states. In the rest of the first BZ, recalling $t(y)<t$, we have $2t(y)\cos(\tfrac{1}{2}k_x\delta_x)<t$ for any $y \in [-\tfrac{W}{2}, \tfrac{W}{2}]$, making the entire GNR unit cell a topological SSH chain possessing a doubly degenerate flat edge state. Therefore, the BZ-wide flat bands are the edge states of the twisted GNR.

Before we leave this section, we briefly discuss the effect of electron-electron interactions which we have ignored so far. It is well known that a repulsive Coulomb interaction can create local ferromagnetic order at the two edges of a strain-free zigzag GNR such that the two edges have opposite magnetization \cite{fujita1996, okada2001, son2006a, son2006b, soriano2012}. The most important consequence on band structure is that a small band gap opens up at the Dirac points \cite{son2006a, son2006b, soriano2012}, which should in general deform the pLLs in Eq.~(\ref{pLL}). However, this band gap is inversely proportional to the ribbon width \cite{son2006a}, and quickly becomes much smaller than the pLL spacing for realistic interaction strengths and wide ribbons we consider here. (An additional gap arises at $k=\tfrac{\pi}{\delta_x}$, but is unlikely to affect the pLLs located within $[-\tfrac{2\pi}{3\delta_x}, \tfrac{2\pi}{3\delta_x}]$.) Therefore, we would expect Eq.~(\ref{pLL}) to characterize the pLL dispersions even in the presence of interaction. Similar to the zigzag edge state, the pLL$_0$ may also carry local ferromagnetic order and form in the twisted zigzag GNR an ``edge-compensated'' antiferromagnetism \cite{roy2014}.

\section{Transport signatures of the twisted graphene nanoribbon}
\label{s3}
In Sec.~\ref{s2}, we have obtained the pLL dispersions [Eq.~(\ref{pLL})] by studying the Bloch Hamiltonian [Eq.~(\ref{H_Bloch})] in the vicinity of $y_0$, where the bulk zero mode, i.e., the pLL$_0$, is centered. In the present section, we adopt these results to study the transport signatures of the twisted GNR at low energies.

\subsection{Density of states}
\label{s3a}
To derive the transport properties of the twisted GNR, it is instructive to first investigate the density of states (DOS) contributed by the doubly degenerate pLLs and the two Dirac cones from which these pLLs emerge. Without loss of generality, we will set the chemical potential $\mu>0$ in the following, while the $\mu<0$ case can be treated using the particle-hole transformation.

We first examine the pLLs whose dispersions have been found in Eq.~(\ref{pLL}). It is critically important to note that these pLLs are only well-defined between the Dirac cones. Therefore, the actual extent of the $n$-th pLL (pLL$_n$) is $[-k_x^n, k_x^n]$, which is only a portion of the domain $[-\tfrac{2\pi}{3\delta_x}, \tfrac{2\pi}{3\delta_x}]$ except when $n=0$ [Fig.~\ref{fig2}(a)]. The bound $k_x^n$ can be determined by finding the intersection of $\epsilon_n^+(k_x)$ with the Dirac cone envelope $\varepsilon_D^+(k_x)= [2\cos(\tfrac{1}{2}k_x\delta_x)-1]t$. Upon finding $k_x^n$, the DOS contributed by pLLs is
\begin{equation} \label{dos_pLL}
\begin{split}
g_L(\mu) &=2 \sum_n \int_{-k_x^n}^{k_x^n} \frac{dk_x}{2\pi} \delta[\mu - \epsilon_n^+(k_x)] = \frac{2}{\pi}\sum_n  \frac{\nu_n(\mu)}{\frac{d\epsilon_n^+}{dk_x}|_{\mu}} ,
\end{split}
\end{equation}
where $\frac{d\epsilon_n^+}{dk_x}|_{\mu}$ is calculated in the left BZ here and below. In Eq.~(\ref{dos_pLL}), we also define for the pLL$_n$ the occupancy parameter $\nu_n(\mu) = \theta(\epsilon_n^\Gamma - \mu) - \theta(\epsilon_n^D - \mu)$ with $\theta(\cdot)$ being the Heaviside step function and $\epsilon_n^D \equiv \epsilon_n^+(-k_x^n) = \varepsilon_D^+(-k_x^n)$ [$\epsilon_n^\Gamma \equiv \epsilon_n^+(0)$] marking the minimum (maximum) of the electron-like pLL$_n$. The values of $\epsilon_n^D$ and $\epsilon_n^\Gamma$ for the lowest few pLLs are listed in Table~\ref{tab1}.

\begin{table}[tb] 
\caption{The energy bounds of lowest 20 pLLs in a twisted GNR with $\lambda=0.0005 a^{-1}$ and $N=1200$.} \label{tab1}
\renewcommand*{\arraystretch}{1.5}

\begin{tabular}{C{0.8cm}C{1.5cm}C{1.5cm}|C{0.8cm}C{1.5cm}C{1.5cm}}
\hline\hline
$n$ & $\epsilon_n^D/t$ & $\epsilon_n^\Gamma/t$ & $n$ & $\epsilon_n^D/t$ & $\epsilon_n^\Gamma/t$ \\
\hline
   
$1$ & $0.0224$ & $0.0496$ &   $11$ & $0.1080$ & $0.1644$\\ 

$2$ & $0.0354$ & $0.0701$ &  $12$ & $0.1143$ & $0.1717$ \\

$3$ & $0.0463$ & $0.0859$ &  $13$ & $0.1203$ & $0.1787$ \\ 

$4$ & $0.0559$ & $0.0991$ &  $14$ & $0.1262$ & $0.1854$ \\  

$5$ & $0.0647$ & $0.1108$ &  $15$ & $0.1320$ & $0.1920$ \\  

$6$ & $0.0729$ & $0.1214$ &  $16$ & $0.1375$ & $0.1983$ \\  

$7$ & $0.0806$ & $0.1311$ &  $17$ & $0.1430$ & $0.2044$ \\ 

$8$ & $0.0879$ & $0.1402$ & $18$ & $0.1483$ & $0.2103$ \\   

$9$ & $0.0949$ & $0.1487$ & $19$ & $0.1536$ & $0.2160$ \\ 

$10$ & $0.1016$ & $0.1567$ &  $20$ & $0.1587$ & $0.2217$  \\
  
\hline\hline
\end{tabular}
\end{table}

\begin{figure}[tb]
\includegraphics[width = 8.6cm]{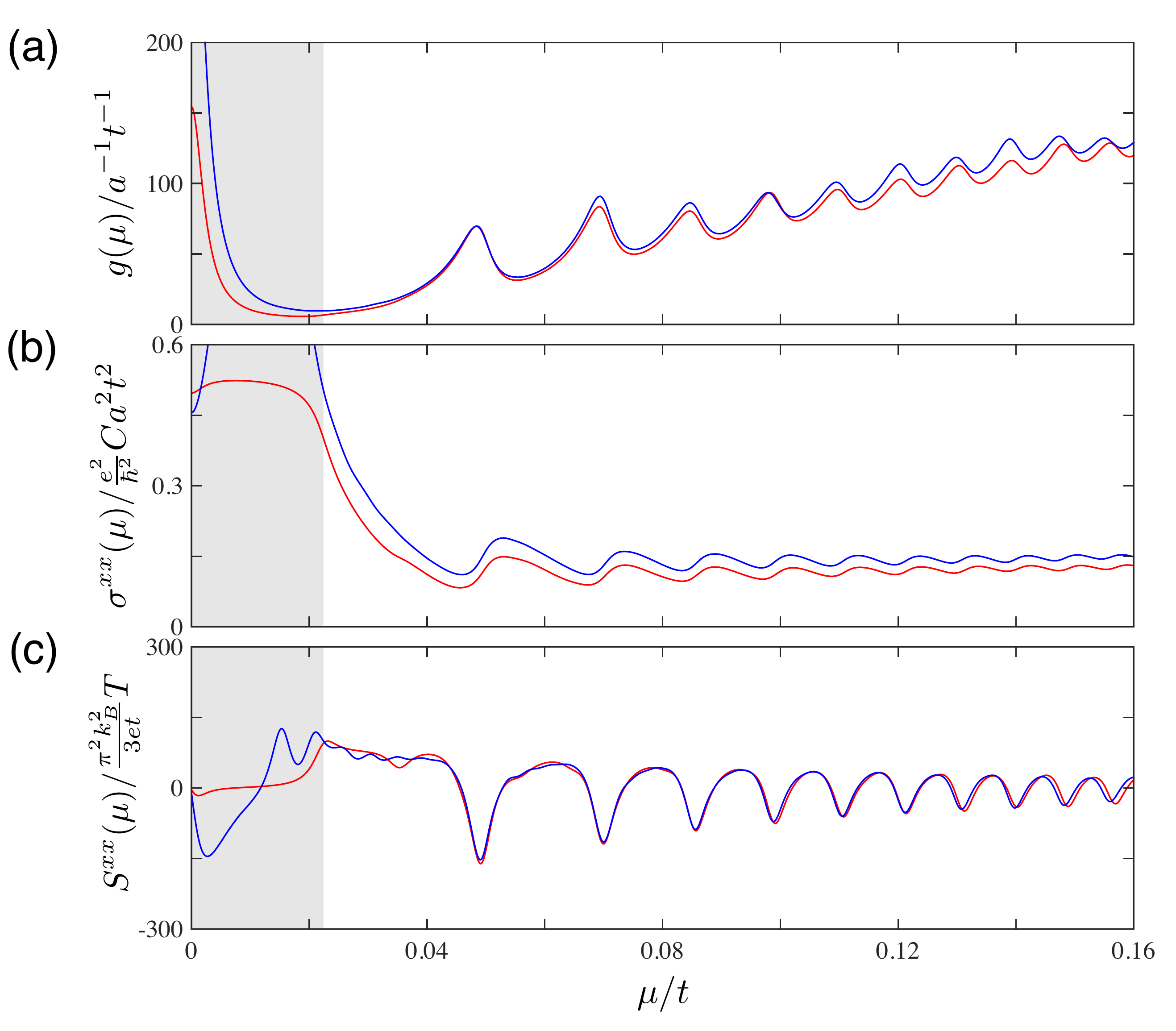}
\caption{Transport properties of the twisted GNR vs the chemical potential $\mu$. Red and blue curves stand for the analytical and numerical results, respectively. Shaded areas denote the only direct band gap of pLLs, which is located between the pLL$_0$ and the pLL$_1$. (a) DOS of the twisted GNR. Note that the mismatch between the numerical DOS and the analytical DOS [Eq.~(\ref{dos})] in the pLL gap is because Eq.~(\ref{dos}) excludes the contribution of the flat edge state [Fig.~\ref{fig2}(a)]. (b) The longitudinal electrical conductivity whose theoretically proposed value is given by Eq.~(\ref{cond}). (c) The Seebeck coefficient calculated from panel (b) using the Mott relation [Eq.~(\ref{seebeck})]. For all panels, the data are broadened by convolving in energy a Lorentzian of width $\delta_\epsilon=0.0024t$ to simulate the effects of disorder and finite temperature. } \label{fig3}
\end{figure}

As for the bulk bands harbored by the Dirac cones, we may treat their contribution to the DOS as identical to that of Dirac cones in a strain-free GNR of renormalized width, as explained in Sec.~\ref{s2} and Figs.~\ref{fig2}(b) and~\ref{fig2}(c). In the limit of large GNR width, the DOS associated with the bulk states in the Dirac cones can then be written as
\begin{equation} \label{dos_DC}
g_D(\mu)= 2 W \xi(\mu) \int\frac{d\bm q}{(2\pi)^2} \delta[\mu - \varepsilon(\bm q)] = \frac{4N}{3\pi}\frac{\mu}{at^2} \xi(\mu),
\end{equation}
where $\varepsilon(\bm q) = \pm \tfrac{3}{2}ta (q_x^2+q_y^2)^{1/2}$ are the dispersions characterizing the Dirac cones before projection into the one-dimensional BZ, and the multiplier $\xi(\mu)=\mathcal{N}_{\lambda}(\mu)/\mathcal{N}_0(\mu)$ concerns the difference in energy band numbers with $\mathcal{N}_{\lambda}(\mu) \approx 1+2 \sum_{n>0} \theta(\mu-\epsilon_n^D)$ [$\mathcal{N}_0(\mu) \approx \tfrac{2N}{\pi t} \mu$] being the number of bands in a single Dirac cone of the twisted (undeformed) GNR intersecting the chemical potential $\mu>0$. A more substantial derivation of $\xi(\mu)$ is provided in Appendix~\ref{a2}. The total DOS is then the combination of Eqs.~(\ref{dos_pLL}) and~(\ref{dos_DC}),
\begin{equation} \label{dos}
g(\mu) =\frac{2}{\pi}\sum_n  \frac{1}{\frac{d\epsilon_n^+}{dk_x}|_{\mu}} \nu_n(\mu) + \frac{4N}{3\pi}\frac{\mu}{at^2} \xi(\mu),
\end{equation}
which is substantiated by comparing to the DOS of the nanoribbon tight-binding Hamiltonian [Eq.~(\ref{H_ribbon})] numerically evaluated through the tetrahedron method \cite{blochl1994} as illustrated in Fig.~\ref{fig3}(a).

%

\subsection{Negative strain-resistivity}
\label{s3b}
We now present the analysis of the longitudinal electrical conductivity of the twisted GNR. At sufficiently low temperatures, we may apply the Sommerfeld expansion to the Boltzmann formalism and keep only the lowest order contribution, which, for the pLLs, reads 
\begin{align} \label{cond_pLL}
\sigma_L^{xx}(\mu) 
&=2e^2 \sum_n \tau_n(\mu) \int_{-k_x^n}^{k_x^n} \frac{dk_x}{2\pi} [v_n^x(k_x)]^2 \delta[\mu-\epsilon_n^+(k_x)] \nonumber
\\
&= \frac{2e^2}{\pi \hbar^2} \tau(\mu) \sum_n \frac{d\epsilon_n^+}{dk_x}\bigg|_\mu \nu_n(\mu),
\end{align}
where $v_n^x(k_x) = \tfrac{1}{\hbar}\frac{d\epsilon_n^+}{dk_x}$ is the band group velocity for the pLL$_n$ and we have used the energy-dependent relaxation time approximation \cite{ashcroft1976} and assumed identical relaxation time $\tau(\mu)$ for all energy bands. The contribution to the electrical conductivity from the bulk states in the Dirac cones may be written as
\begin{align} \label{cond_DC}
\sigma_D^{xx}(\mu) &= 2e^2W \chi(\mu) \int \frac{d\bm q}{(2\pi)^2} \tau(\mu) v_x^2(\bm q)  \delta[\mu-\varepsilon(\bm q)] \nonumber
\\
&= \frac{3N}{2\pi}  \chi(\mu) \frac{e^2}{\hbar^2} \tau(\mu) a\mu,
\end{align}
where $v_x(\bm q) = \tfrac{1}{\hbar} \tfrac{\partial \varepsilon(\bm q)}{\partial q_x}$ is the band velocity associated with the Dirac cone. Similar to the DOS contribution [Eq.~(\ref{dos_DC})], which is modified from the strain-free Dirac cone contribution by a multiplier $\xi(\mu)$, the conductivity contribution [Eq.~(\ref{cond_DC})] also requires a multiplier $\chi(\mu)$ to incorporate the difference in band numbers. We argue that $\chi(\mu) \approx \tfrac{1}{2}\xi(\mu)$ because the energy bands in Dirac cones of the twisted GNR are no longer ``V-shaped'' as those hosted by the strain-free Dirac cones, but are nearly half ``V-shaped'' as illustrated in Fig.~\ref{fig2}(b), making the number of Dirac cones contributing to the electrical conductivity effectively \emph{one} (see Appendix~\ref{a2}). 

A comprehensive understanding of the electrical conductivity also requires the knowledge of the relaxation time $\tau(\mu)$, especially its energy dependence, which is sensitive to the details of scattering mechanism. For strain-free graphene, the screened Coulomb scattering dominates \cite{hwang2009a, hwang2007a} and the relaxation time becomes $\tau(\mu) \propto \mu$ due to the ``Diracness'' of the charge carriers \cite{hwang2009b, hwang2007b}. In the presence of twist, the Dirac cones are broken into pLLs and the $\mu$ dependence of the relaxation time would change. Using Fermi's golden rule \cite{zhangjm2016}, the scattering time can be written as $\tau(\mu)=C/g(\mu)$, where $C= \tfrac{\hbar}{2\pi} |V_{if}|^{-2}$ encodes the detailed information of the scattering mechanism with $V_{if}$ being the scattering matrix element between the initial state $\ket i$ and the final state $\ket f$. When the Drude contribution is the major source of scattering, the parameter $C$ becomes a constant proportional to the density of the impurities \cite{lantagne2020} consistent with the first-order Born approximation prediction \cite{doniach1998}. For other types of scatterers, such as charged impurities, the parameter $C$ is expected to have a smooth dependence on $\mu$. In terms of the parameter $C$, the total longitudinal electrical conductivity comprising Eqs.~(\ref{cond_pLL}) and (\ref{cond_DC}) reads
\begin{equation} \label{cond}
\sigma^{xx}(\mu) =\frac{\frac{2}{\pi} \sum_n \frac{1}{at}\frac{d\epsilon_n^
+}{dk_x}|_\mu \nu_n(\mu) + \frac{3N}{2\pi} \frac{\mu}{t} \chi(\mu) } {\frac{2}{\pi}\sum_n  \frac{1}{\frac{1}{at} \frac{d\epsilon_n^+}{dk_x}|_{\mu}} \nu_n(\mu) + \frac{4N}{3\pi}\frac{\mu}{t} \xi(\mu)} \frac{e^2}{ \hbar^2} C a^2t^2.
\end{equation}
Assuming Drude scattering, the electrical conductivity $\sigma^{xx}$ is calculated as a function of $\mu$ in Fig.~\ref{fig3}(b); our theoretical prediction Eq.~(\ref{cond}) is seen to capture the essential features of the numerical results. Though Fig.~\ref{fig3}(b) is quantitatively correct only for $C=\text{const.}$, the pLLs should generally manifest themselves through singularities in the DOS $g(\mu)$ and hence dips in $\sigma^{xx}$, as long as $C$ has a smooth $\mu$ dependence.

\begin{figure}[tb]
\includegraphics[width = 8.6cm]{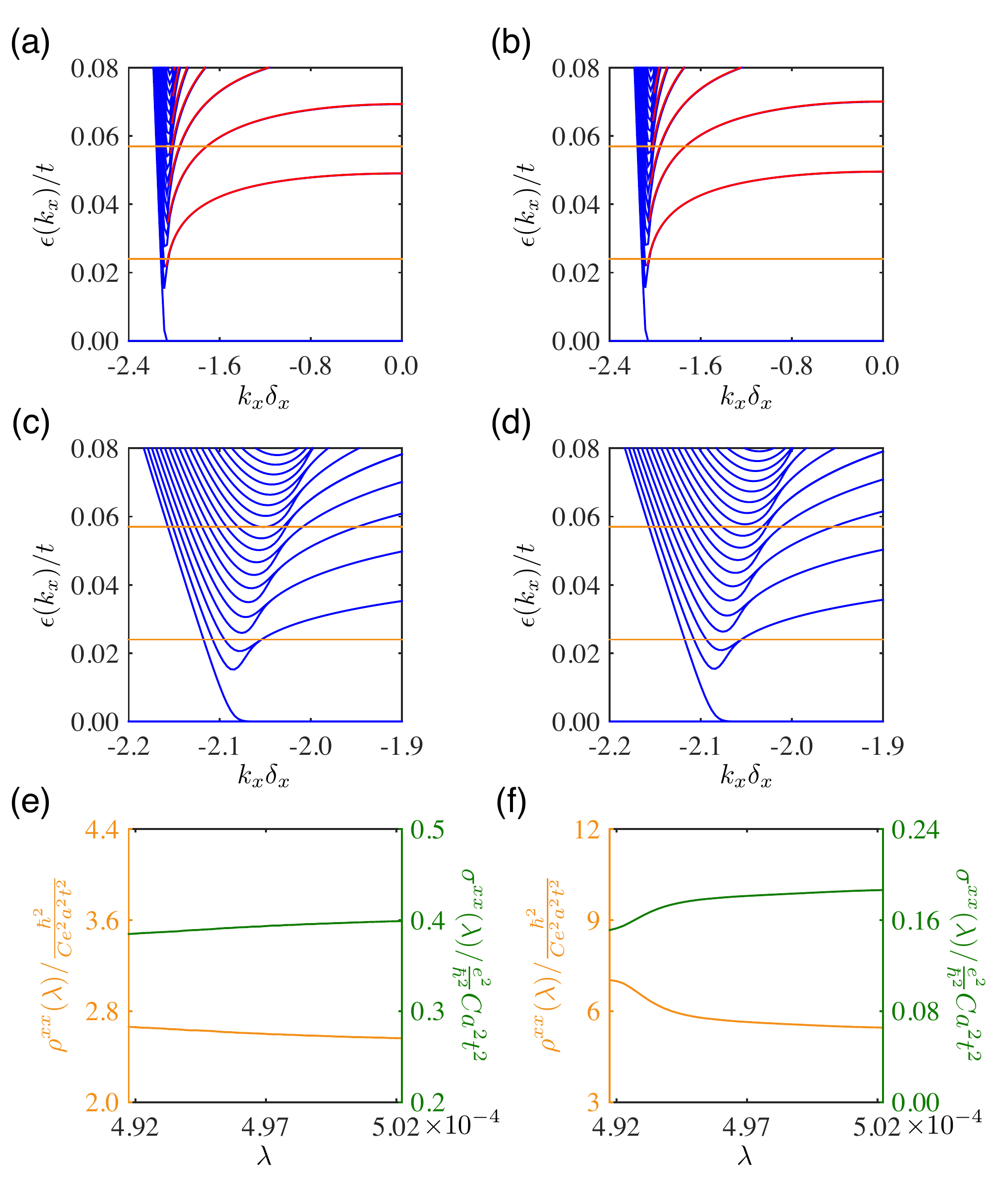}
\caption{Strain resistivity of a twisted GNR with $N=1200$. [(a),(b)] Numerically calculated energy bands (blue) and analytically predicted pLLs [Eq.~(\ref{pLL})] (red) for (a) $\lambda a=4.917\times 10^{-4}$ and (b) $\lambda a=5.022\times 10^{-4}$. [(c),(d)] Enlargements of panels (a) and (b) in the vicinity of the Dirac point $K$. (e)--(f) Resistivity (orange) and conductivity (green) as functions of the twist $\lambda$ at (e) $\mu=0.024t$ and (f) $\mu=0.057t$. These two chemical potentials are marked by horizontal orange lines in panels (a)--(d). The resistivity and conductivity data are broadened by convolving in energy a Lorentzian of width $\delta_\epsilon=10^{-4}t$.} \label{fig4}
\end{figure}

We now analyze the $\lambda$ dependence of the electrical conductivity [Eq.~(\ref{cond})]. When the twist $\lambda$ is allowed to vary in a narrow range in which only the dispersive pLL$_1$ is partially filled with the chemical potential $\mu_1=0.024t$ [lower orange lines, Figs.~\ref{fig4}(a) to~\ref{fig4}(d)], the values of the occupancy parameters $\nu_n(\mu)$ and the multipliers $\xi(\mu)$ and $\chi(\mu)$ remain intact. Moreover, the band velocities $\sim\tfrac{d\epsilon_n}{dk_x}$ of the pLLs are increasing functions of $\lambda$ (see Appendix~\ref{a1}). Consequently, both the total electrical conductivity $\sigma^{xx}$ and the pLL contribution $\sigma_L^{xx}$ increase with the twist $\lambda$, giving rise to a negative strain-resistivity $\rho^{xx} = 1/\sigma^{xx}$ \footnote{Because the twist lattice deformation does not break the time-reversal symmetry, the Hall conductance $\sigma^{xy}$ must vanish, giving rise to a diagonal conductivity tensor $\bm \sigma = \text{diag}(\sigma^{xx}, \sigma^{yy})$. The resistivity tensor is then $\bm \rho = \bm \sigma^{-1} = \text{diag}(1/\sigma^{xx}, 1/\sigma^{yy})$.} as illustrated by the orange curve in Fig.~\ref{fig4}(e). Such an effect is analogous to the negative magnetoresistivity \cite{huang2015, kim2013, xiong2015, zhang2016, son2013, burkov2015, li2016} in topological semimetals with only chiral LLs (cf. pLL$_1$) partially filled (i.e. in the quantum limit) and may serve as a manifestation of the $(1+1)$-dimensional chiral anomaly \cite{nielsen1983}, which coincides with the valley anomaly in graphene \cite{lantagne2020}. For a higher chemical potential $\mu_2=0.057t$ [upper orange line, Figs.~\ref{fig4}(a) to~\ref{fig4}(d)] outside the quantum limit, pLL$_1$ (pLL$_{2,3,4}$) is always fully (partially) filled during the variation of the twist; a negative strain resistivity arising from pLL$_{2,3,4}$ then emerges [Fig.~\ref{fig4}(f)]. We mention that the pLL-induced negative strain resistivity can only occur when $\lambda$ is scanned in a narrow range; otherwise the bottoms of subbands in the Dirac cones [Figs.~\ref{fig4}(c) and~\ref{fig4}(d)] can in general cross the chemical potential, giving rise to quantum-oscillation-like signals in both the resistivity and the conductivity. Consequently, the negative strain resistivity may become obscured. Lastly, we note that the thermal conductivity of the GNR is related to the electrical conductivity through the Wiedemann-Franz law $\kappa^{xx}=\tfrac{\pi^2 k_B^2}{3e^2}T \sigma^{xx}$. Therefore, it also increases with $\lambda$ in the quantum limit, giving rise to a negative strain-thermal resistivity similar to that in Weyl superconductors \cite{kobayashi2018}.

\subsection{Enhanced thermopower} 
\label{s3c}
We now study the thermoelectric effect in the twisted GNR. Because of the preservation of the time-reversal symmetry, the twisted GNR exhibits no Nernst effects or thermal Hall effects in the presence of a temperature gradient $\partial_x T$. However, the Seebeck coefficient is compatible with the time-reversal symmetry and can be conveniently determined through the Mott relation \cite{cutler1969}
\begin{equation} \label{seebeck}
S^{xx}(\mu)= - \frac{\pi^2 k_B^2 T}{3e\sigma^{xx}} \frac{d\sigma^{xx}}{d\mu},
\end{equation}
which arises from the Sommerfeld expansion \cite{ashcroft1976} at low temperatures $k_BT \ll \mu$. For the broadening $\delta_\epsilon=0.0024t$ chosen for $\sigma^{xx}$ in Fig.~\ref{fig3}(b), the temperature is $k_BT \sim \delta_\epsilon$. If we are only interested in pLLs (outside the gray patches in Fig.~\ref{fig3}) which are far from the charge neutrality point, $k_BT \ll \mu$ indeed stands and the application of the Mott relation is legitimate. On the other hand, for higher temperatures comparable to the Fermi temperature, a deviation from the Mott relation has been observed \cite{checkelsky2009, zuev2009, wei2009}. The thermopower becomes sensitively dependent on the scattering mechanisms \cite{hwang2009a, hwang2007a, ghahari2016, xie2016} and is likely to be dominated by charged impurities \cite{hwang2009a, hwang2007a}, while electron-electron scattering and electron-optical-phonon scattering may also contribute \cite{ghahari2016, xie2016}.

We note that every time a pLL is fully filled, $\sigma^{xx}$ is greatly suppressed but experiences a sudden boost giving rise to large values of both $\frac{1}{\sigma^{xx}}$ and $|\frac{d\sigma^{xx}}{d\mu}|$, thus significantly enhancing the Seebeck coefficient as illustrated by the sequence of dips around $\epsilon_n^\Gamma$ in Fig.~\ref{fig3}(c), which mimic those produced by the ordinary flat Dirac-Landau levels resulting from uniform real magnetic fields \cite{xing2009, liu2012, checkelsky2009, zuev2009, wei2009}. The enhanced Seebeck coefficient is accompanied by a boosted Peltier coefficient through the Thomson relation $\Pi^{xx}=TS^{xx}$ \cite{ashcroft1976}. 

\section{Discussions and Conclusions}
\label{s4}
During our derivation of the pLL dispersions and the associated transport properties, two assumptions have been made. First, we assume the continuum limit and strongly localized pLLs so that both the shift operators $\hat s_{\pm \delta_y}$ and the hopping $t(y)$ can be expanded only to the linear order. We argue that this is a legitimate estimate when the GNR width is not too small. In fact, as we will show in Appendix~\ref{a3}, the quadratic and cubic order terms in the expansions of $\hat s_{\pm \delta_y}$ and $t(y)$ can only contribute a correction of order of $10^{-4}t$ when $N=1200$, which should be safely negligible. Second, we assume that the effect of the next nearest neighbor hopping is negligible. This is because such an effect can be in principle canceled by a fine-tuned electric field \cite{lantagne2020}. In realistic graphene, the strain-free next nearest neighbor hopping ranges from $0.02t$ to $0.2t$ \cite{neto2009} and thus should deform the twist-induced pLLs to some extent in the absence of the applied electric field. The response of pLLs to the next nearest neighbor hopping is discussed in detail in Appendix~\ref{a4}.

To summarize, we have studied the strain-induced pLLs in a twisted GNR. By tracking the formation of the bulk zero mode located on the domain wall of the GNR unit cell, which is effectively an SSH model, we reduce the low-energy theory of the twisted GNR into an exactly solvable minimally coupled Dirac theory to derive the momentum dependence of the pLLs. Such dispersive pLLs produce a negative strain resistivity if partially filled and can enhance the thermopower if fully filled.

The method we have adopted deriving the pLL dispersions is by no means \emph{ad hoc} and should be in principle transplantable to GNRs under other inhomogeneous strain patterns or magnetic fields with complicated or even arbitrary spatial profiles. It can also be conveniently generalized to other graphene-like Dirac materials made of photons \cite{rechtsman2013}, magnons \cite{ferreiros2018}, phonons \cite{wen2019, brendel2017}, and Majorana particles \cite{rachel2016, perreault2017}, since strain-induced flat Dirac-Landau levels have been reported in these materials. Our work may also be adapted to superconducting Dirac matter such as Weyl superconductors \cite{kobayashi2018, liu2017b, matsushita2018} and $d$-wave superconductors \cite{massarelli2017, nica2018}, where strain may be the only hope to Landau-quantize the Bogoliubov quasiparticles.

\begin{acknowledgments}
The authors are indebted to R. Moessner, M. Franz, P. W. Brouwer, E. Lantagne-Hurtubise, X. -X. Zhang and L. He for insightful discussions. Z.S. is supported in part by project A02 of the CRC-TR 183. H.-Z.L. is supported by the National Natural Science Foundation of China (Grants No~11534001, No.~11974249, and No.~11925402), the National Basic Research Program of China (Grant No.~2015CB921102), the Strategic Priority Research Program of Chinese Academy of Sciences (Grant No.~XDB28000000), Guangdong Province (Grants No.~2016ZT06D348 and No.~2020KCXTD001), the National Key R \& D Program (Grant No.~2016YFA0301700), Shenzhen High-level Special Fund (Grants No.~G02206304 and No.~G02206404), and the Science, Technology and Innovation Commission of Shenzhen Municipality (Grants No.~ZDSYS20170303165926217, No.~JCYJ20170412152620376, and No.~KYTDPT20181011104202253).  
\end{acknowledgments}

\appendix
\section{Derivation of pLL dispersions}
\label{a1}
In Sec.~\ref{s2} of the main text, we have derived the dispersions [Eq.~(\ref{pLL})] of the pLLs in a twisted GNR. We here present a detailed solution of the eigenvalue problem of the minimally coupled Dirac Hamiltonian $\mathcal H_{k_x,y}$ [Eq.~(\ref{H_Dirac})], from which the pLLs arise. Explicitly, the eigenvalue problem of $\mathcal H_{k_x,y}$ is written as
\begin{equation} \label{Dirac_eqs}
\begin{split}
\bigg[ \Omega_{y_0}(y-y_0) - t\delta_y \frac{d}{dy} \bigg] f_-(y) = \epsilon f_+(y),
\\
\bigg[ \Omega_{y_0}(y-y_0) + t\delta_y  \frac{d}{dy} \bigg] f_+(y) = \epsilon f_-(y),
\end{split}
\end{equation}
where $f_+(y)$ and $f_-(y)$ constitute the eigenvector $\Psi(y) = e^{ik_xx}  [f_+(y), f_-(y)]^T$ corresponding to the eigenenergy $\epsilon$. In a more compact form, Eq.~(\ref{Dirac_eqs}) can be rewritten as
\begin{equation} \label{Dirac_eq}
\bigg[ \Omega_{y_0}^2(y-y_0)^2 - t^2 \delta_y^2 \frac{d^2}{dy^2} - s t \delta_y \Omega_{y_0} \bigg] f_s(y) =\epsilon^2 f_s(y),
\end{equation}
with $s=\pm$. For transparency, we introduce the dimensionless coordinate
\begin{equation} \label{xi}
\xi_y=\sqrt{\bigg|\frac{\Omega_{y_0}}{t\delta_y}\bigg|} (y-y_0),
\end{equation}
which helps simplify Eq.~(\ref{Dirac_eq}) as
\begin{equation} \label{Dirac_eq_simp}
 \frac{d^2 f_s}{d\xi_y^2}-\xi_y^2 f_s+\bigg[\frac{\epsilon^2}{| t \delta_y \Omega_{y_0} |} + s \cdot \text{sgn}(t \delta_y \Omega_{y_0}) \bigg] f_s = 0. 
\end{equation}
We define an auxiliary function $g_s(\xi_y)=e^{\xi_y^2/2}f_s(\xi_y)$ that transforms Eq.~(\ref{Dirac_eq_simp}) into the exactly solvable Hermite's differential equation
\begin{equation} \label{Hermite}
\frac{d^2g_s}{d \xi_y^2} - 2\xi_y \frac{d g_s}{d \xi_y} + \bigg[\frac{\epsilon^2}{|t \delta_y \Omega_{y_0} |} + s \cdot \text{sgn} (t \delta_y \Omega_{y_0}) -1 \bigg] g_s=0,
\end{equation}
which possesses square-integrable solutions only when the eigenenergy $\epsilon$ adopts discrete values
\begin{equation} \label{epsilon_n}
\epsilon_n^2 = 2n |t \delta_y \Omega_{y_0} |,
\end{equation}
where $n$ is a nonnegative integer. The solutions of Eq.~(\ref{Hermite}) are the Hermite polynomials $g_{\text{sgn}(t\delta_y\Omega_{y_0})}(\xi_y)=H_n(\xi_y)$ and $g_{-\text{sgn}(t\delta_y\Omega_{y_0})}(\xi_y)= \text{sgn}(n) H_{n-1}(\xi_y)$, making the eigenvectors 
\begin{equation} \label{eigenvector}
\begin{split}
\Psi_{n>0}^\pm(y) &= \frac{1}{\sqrt{2^{n+1} \pi^{\frac{1}{2}} n!}} e^{ik_xx} e^{-\frac{1}{2} \xi_y^2} 
\begin{bmatrix}
\pm  H_{n}(\xi_y)
\\
\sqrt{2n} H_{n-1}(\xi_y)
\end{bmatrix}, 
\\
\Psi_0(y) &= \frac{1}{\sqrt{\pi^{\frac{1}{2}}}} e^{ik_xx} e^{-\frac{1}{2} \xi_y^2} 
\begin{bmatrix}
H_0(\xi_y)
\\
0
\end{bmatrix},
\end{split}
\end{equation}
when $\text{sgn}(t\delta_y\Omega_{y_0})>0$. When $\text{sgn}(t\delta_y\Omega_{y_0})<0$, the eigenvectors should be written as $\tau^x \Psi_{n>0}^\pm(y)$ and $\tau^x \Psi_0(y)$. The $y$-dependence of the eigenvectors is incorporated through Eq.~(\ref{xi}). By plugging these acquired eigenvectors into Eq.~(\ref{Dirac_eqs}), we may specify the sign of the eigenenergy $\epsilon_n$, which is not encoded in Eq.~(\ref{epsilon_n}).  Specifically, the eigenenergies associated with the eigenvectors $\Psi_{n>0}^\pm(y)$ [$\tau^x \Psi_{n>0}^\pm(y)$] and $\Psi_0(y)$ [$\tau^x \Psi_0(y)$] are $\epsilon_{n>0}^\pm=\pm \sqrt{2n|t\delta_y \Omega_{y_0}|}$ ($\epsilon_{n>0}^\pm=\mp \sqrt{2n|t\delta_y \Omega_{y_0}|}$)  and $\epsilon_0^+=0$ ($\epsilon_0^-=0$), respectively. The pLL dispersions can then be written as Eq.~(\ref{pLL}), which possesses two-fold degeneracy.

As analyzed in Sec.~\ref{s3}, the dispersive pLLs impact the transport properties, e.g., Eqs.~(\ref{dos}) and (\ref{cond}), through the derivative of Eq.~(\ref{pLL}) calculated at the left BZ. Explicitly, it reads
\begin{equation} \label{derivative}
\frac{d\epsilon_n^+}{dk_x}\bigg|_\mu = \frac{3 \sqrt 3}{4 g} \frac{\sqrt{ \tfrac{2}{\sqrt 3} n g \lambda a}}{[\gamma_\mu^2(\gamma_\mu^2-1)]^{3/4}} \frac{\sqrt{1-\tfrac{1}{4} e^{2g(\gamma_\mu-1)} }} {e^{g(\gamma_\mu-1)}} at,
\end{equation}
where we have defined the parameter 
\begin{equation}
\gamma_\mu=\frac{1}{\sqrt{1-4\mu^4/27t^4 n^2 g^2 \lambda^2 a^2}}.
\end{equation}
It is straightforward to find that $\frac{d\epsilon_n^+}{dk_x}|_\mu$ is an increasing function of the twist $\lambda$ because $\gamma_\mu$ is a decreasing function of $\lambda$. Therefore, if $\lambda$ is allowed to vary in a narrow range such that only pLL$_1$ is partially filled, the electrical conductivity  Eq.~(\ref{cond}) becomes an increasing function of $\lambda$, resulting in a negative strain resistivity.

\section{Derivation of multipliers}
\label{a2}
The fact that Dirac cones of a twisted GNR [Fig.~\ref{fig2}(b)] contain far fewer bands than those of an undeformed GNR [Fig.~\ref{fig2}(c)] requires multipliers $\xi(\mu)$ and $\chi(\mu)$ to be introduced when calculating the DOS [Eq.~(\ref{dos_DC})] and the electrical conductivity [Eq.~(\ref{cond_DC})], respectively. To figure out the values of these multipliers, we now analyze the bulk bands hosted by the left Dirac cone (i.e., the one located at $k_D=-\tfrac{2\pi}{3\delta_x}$) of twisted and/or undeformed GNR with close attention paid to the morphology and transport properties associated with these bulk bands.

\begin{figure} [tb]
\includegraphics[width = 8.6cm]{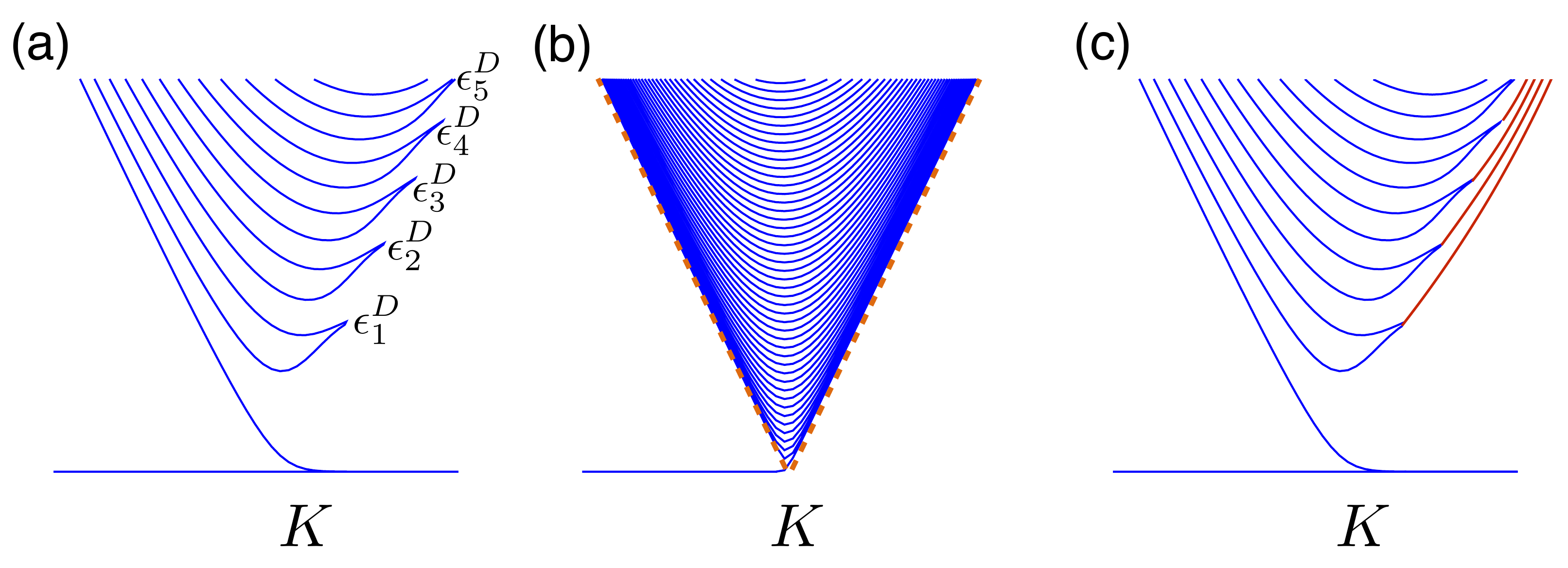}
\caption{Schematic plot of the band structure of the left Dirac cone. (a) The energy bands, labeled as $\varepsilon_n^t(k_x)$,  of a twisted GNR with $\epsilon_{n=1,\cdots,5}^D$ marking the confluence of these bands into pLLs. (b) A Dirac cone of an undeformed GNR. The dashed curve is the boundary of the Dirac cone. The left (right) boundary is characterized by $\varepsilon_D^-(k_x) = t-2t \cos (\tfrac{1}{2} k_x\delta_x)|_{k_x<k_D}$ [$\varepsilon_D^+(k_x) = -t+2t \cos (\tfrac{1}{2} k_x\delta_x)|_{k_x>k_D} $]. (c) An artificial band structure is constructed by adding to (a) energy bands [red and denoted as $\varepsilon_n^a(k_x)]$, which are (almost) parallel to the right boundary of the Dirac cone, making the modified energy bands $\varepsilon_n^i(k_x) = \{\varepsilon_n^t(k_x), \varepsilon_n^a(k_x)\}$ also V shaped, resembling those in panel (b).} \label{fig5}
\end{figure}

For the bulk bands inside the Dirac cone of a twisted GNR, as illustrated in Fig.~\ref{fig5}(a), when $k_x$ increases, an energy band first goes downhill along the left boundary of the Dirac cone and reaches its minimum in the vicinity of the right boundary of the cone. It then climbs uphill toward the right boundary, at which it merges with another energy band and they evolve into a doubly degenerate pLL. The transport associated with this Dirac cone is only contributed by the partially filled energy bands, labeled by $\varepsilon_n^t(k_x)$, which intersect the chemical potential $\mu>0$. The number of such bands, denoted as $\mathcal N_\lambda(\mu)$, can be estimated by counting the number of pLLs emerging from the Dirac cone below the chemical potential $\mu$. Explicitly, we have
\begin{equation} \label{NL}
\mathcal N_\lambda(\mu) \approx 1+2\sum_{n > 0} \theta(\mu-\epsilon_n^D),
\end{equation}
where $\theta(\cdot)$ is the Heaviside step function. The first term in Eq.~(\ref{NL}) relates to the electron-like copy of the doubly degenerate pLL$_0$. The DOS at chemical potential $\mu$ is then
\begin{equation} \label{dos_t}
g_D^t(\mu) = \sideset{}{'}\sum_n  \int \frac{dk_x}{2\pi}  \delta[\mu-\varepsilon_n^t(k_x)] = \frac{1}{2\pi} \sideset{}{'}\sum_n  \frac{1}{\big|\frac{d\varepsilon_n^t}{dk_x}\big|_\mu}, 
\end{equation}
where the primed summation only runs over the $\mathcal N_\lambda(\mu)$ partially filled energy bands. It is worth noting that these energy bands are generally not monotonic and each of them may intersect the chemical potential $\mu$ at most twice. In that case, we may divide $\varepsilon_n^t(k_x)$ at its minimum into two monotonic segments and let the summation run over both segments. Nevertheless, we conclude from Eq.~(\ref{dos_t}) that the DOS is dominated by the energy bands with small group velocities. In contrast, the electrical conductivity obtained through the Boltzmann formalism,
\begin{align} \label{cond_t}
\sigma_D^t(\mu) &=  e^2 \tau(\mu) \sideset{}{'}\sum_n \int \frac{dk_x}{2\pi} \bigg( \frac{1}{\hbar} \frac{d\varepsilon_n^t}{dk_x} \bigg)^2  \delta[\mu-\varepsilon_n^t(k_x)] \nonumber
\\
& = \frac{e^2 \tau(\mu)}{2\pi \hbar^2} \sideset{}{'} \sum_n \Big| \frac{d\varepsilon_n^t}{dk_x} \Big|_\mu,
\end{align}
is mainly contributed by the energy bands with large group velocities.

Inside the Dirac cone of an undeformed GNR, we see that most of the energy bands are V shaped and go downhill (uphill) along the left (right) boundary of the cone, leaving the band minima in the center around $k_D$. In addition, there is a state whose energy is a monotonously increasing function of $k_x$. It evolves into the zigzag edge state away from $k_D$ on the left [Fig.~\ref{fig5}(b)]. The proximity of band minima to the Dirac point $k_D$ indicates a convenient way to estimate the number of partially filled energy bands, denoted as $\mathcal N_0(\mu)$. Specifically, the nanoribbon tight-binding Hamiltonian [Eq.~(\ref{H_ribbon})] is reduced to a Toeplitz tridiagonal matrix at $k_D$, whose eigenvalues are exactly solvable as $\varepsilon_n^u(k_D)=-2t \cos ( \frac{n\pi}{4N+1} )$ with $n=1, 2, \cdots, 4N$. By confining these band minima below the chemical potential $\mu$ but above the Dirac point, we obtain a requirement for band index $n$ as $2N+1 \leq n \leq  \lfloor \tfrac{4N+1}{\pi} \arccos (-\tfrac{\mu}{2t} )  \rfloor$, which is valid if $\mu \geq  -2t \cos (\tfrac{2N+1}{4N+1}\pi)$. For $\mu <-2t \cos (\tfrac{2N+1}{4N+1}\pi)$, the only partially filled energy band is the one evolving into the zigzag edge state, with index $n=2N+1$. This implies the number of energy bands intersecting the chemical potential $\mu$ may be estimated as
\begin{equation} \label{N0}
\begin{split}
\mathcal N_0 (\mu) \approx \bigg \lfloor \frac{4N+1}{\pi} \arccos \bigg(-\frac{\mu}{2t} \bigg) \bigg \rfloor - 2N 
\\
+ \theta \bigg[-\mu -2t \cos \bigg(\frac{2N+1}{4N+1}\pi \bigg) \bigg].
\end{split}
\end{equation}
We note that $\mathcal N_0 (\mu)$ can be greatly simplified if the chemical potential $\mu$ is close to the Dirac point, imposing requirement $\varepsilon_n^u(k_D) \ll t$, or, equivalently, $n-2N \ll N$, on the partially filled energy bands. Such a requirement leads to a uniform subband gap $\Delta \varepsilon_n^u = \varepsilon_{n+1}^u(k_D)-\varepsilon_n^u(k_D) \approx \tfrac{\pi t}{2N}$, which gives an alternative estimate $\mathcal N_0 (\mu) \approx \frac{\mu}{\Delta \varepsilon_n^u} = \frac{2N}{\pi t} \mu$. In the limit of large GNR width, these $\mathcal N_0 (\mu)$ energy bands contribute to the DOS
\begin{equation} \label{dos_u}
g_D^u(\mu) = W \int \frac{d\bm q}{(2\pi)^2} \delta[\mu-\varepsilon(\bm q)] = \frac{2N\mu}{3\pi at^2} = \frac{1}{2\hbar v_F} \mathcal N_0(\mu), 
\end{equation} 
where $v_F=\tfrac{3at}{2\hbar}$ is the Fermi velocity of the Dirac cone. As for the electrical conductivity, we have
\begin{align} \label{cond_u}
\sigma_D^u(\mu) &= e^2 W \tau(\mu) \int \frac{d\bm q}{(2\pi)^2}  \bigg[\frac{1}{\hbar} \frac{\partial \varepsilon(\bm q)}{dq_x} \bigg]^2 \delta[\mu-\varepsilon(\bm q)] \nonumber
\\
&= \frac{3N}{4\pi} \frac{e^2}{\hbar^2} \tau(\mu) a\mu = \frac{e^2 v_F}{4\hbar} \tau(\mu) \mathcal N_0(\mu),
\end{align}
which also depends on $v_F$ and $\mathcal N_0(\mu)$.

To figure out the values of the multipliers, a connection between $g_D^t(\mu)$ [$\sigma_D^t(\mu)$] and $g_D^u(\mu)$ [$\sigma_D^u(\mu)$] is useful. Such a connection can be established by introducing an artificial band structure, in which we require the energy bands $\varepsilon_n^t(k_x)$ to merge with the right boundary of the Dirac cone rather than penetrating the cone and becoming pLLs. Equivalently, each energy band $\varepsilon_n^t(k_x)$ is spliced with an artificial segment $\varepsilon_n^a(k_x)$, whose group velocity is $\sim \tfrac{3at}{2\hbar}$, as illustrated in Fig.~\ref{fig5}(c). The reshaped energy bands, labeled as $\varepsilon_n^i(k_x) = \{\varepsilon_n^t(k_x), \varepsilon_n^a(k_x)\}$ are also V-shaped, resembling those in Fig.~\ref{fig5}(b). Therefore, we may treat the artificial band structure in Fig.~\ref{fig5}(c) as a Dirac cone in an undeformed GNR whose effective width is smaller than the one considered in Fig.~\ref{fig5}(b), because the subband gap is larger. Since the Fermi velocity associated with this Dirac cone is approximately $v_F=\tfrac{3at}{2\hbar}$, by comparing to Eqs.~(\ref{dos_u}) and (\ref{cond_u}), the associated DOS $g_D^i(\mu)$ and electrical conductivity $\sigma_D^i(\mu)$ may be written down as
\begin{subequations} \label{iu}
\begin{align}
g_D^i(\mu)=g_D^u(\mu) \frac{\mathcal N_\lambda(\mu)}{\mathcal N_0(\mu)},
\\ 
\sigma_D^i(\mu) = \sigma_D^u(\mu) \frac{\mathcal N_\lambda(\mu)}{\mathcal N_0(\mu)}.
\end{align}
\end{subequations}
Alternatively, by comparing to Eqs.~(\ref{dos_t}) and (\ref{cond_t}), we write $g_D^i(\mu)$ and $\sigma_D^i(\mu)$ as
\begin{subequations} \label{it}
\begin{align}
g_D^i(\mu) &= g_D^t(\mu)+ \frac{1}{2\pi} \sideset{}{'}\sum_n  \frac{1}{\big|\frac{d\varepsilon_n^a}{dk_x}\big|_\mu} \approx g_D^t(\mu),
\\
\sigma_D^i(\mu) &= \sigma_D^t(\mu)+\frac{e^2 \tau(\mu)}{2\pi \hbar^2} \sideset{}{'} \sum_n \Big| \frac{d\varepsilon_n^a}{dk_x} \Big|_\mu \approx 2 \sigma_D^t(\mu).
\end{align}
\end{subequations}
where we have noticed that the group velocities ($\sim \tfrac{3at}{2\hbar}$) of the artificially added energy bands $\varepsilon_n^a(k_x)$ are much larger than those associated with the band bottoms of $\varepsilon_n^t(k_x)$, whose contributions are enclosed in $g_D^t(\mu)$, so that the dominating contribution to $g_D^i(\mu)$ is from $g_D^t(\mu)$. On the other hand, the large band velocities indicate that the artificially added energy bands $\varepsilon_n^a(k_x)$ should have a significant contribution to the electrical conductivity $\sigma_D^i(\mu)$. The contribution from $\varepsilon_n^a(k_x)$ should be roughly the same as that from the energy bands $\varepsilon_n^t(k_x)$, because the band bottoms of $\varepsilon_n^t(k_x)$ barely affect the transport in the Boltzmann formalism.

We make use of Eqs.~(\ref{iu}) and (\ref{it}) and obtain the multipliers as
\begin{subequations} \label{multipliers}
\begin{align}
\xi(\mu) &=\frac{g_D^t(\mu)}{g_D^u(\mu)} \approx \frac{g_D^i(\mu)}{g_D^u(\mu)} = \frac{\mathcal N_\lambda(\mu)}{\mathcal N_0(\mu)},
\\
\chi(\mu) &= \frac{\sigma_D^t(\mu)}{\sigma_D^u(\mu)} \approx \frac{1}{2}\frac{\sigma_D^i(\mu)}{\sigma_D^u(\mu)} = \frac{1}{2} \frac{\mathcal N_\lambda(\mu)}{\mathcal N_0(\mu)},
\end{align}
\end{subequations}
which result in a transport theory [Eqs.~(\ref{dos}) and (\ref{cond})] that can actually capture the most important features of the numerical results.

\section{Higher order terms in the continuum theory}
\label{a3}
The pLL dispersions in the main text [Eq.~(\ref{pLL})], while in good agreement with the numerics [Fig.~\ref{fig2}(a)], are obtained by expanding both the modified hopping $t(y)$ and the shift operators $\hat{s}_{\pm \delta_y}$ to the linear order in the nanoribbon tight-binding Hamiltonian [Eq.~(\ref{H_ribbon})]. This approximation becomes accurate under the following conditions: (i) the wave functions vary slowly on the lattice scale $\delta_y$, such that the linearization of $\hat{s}_{\pm \delta_y}$ is legitimate; and (ii) the bulk zero mode is strongly localized at the domain wall $y_0$ such that the widths of the pLL wave functions are sufficiently small, making the linear expansion of $t(y)$ valid. Such a requirement also guarantees minimal overlap between the pLLs and the edge states. In more realistic conditions, we also need to consider in the expansions the higher order terms, which turn out to account for the slight discrepancy between Eq.~(\ref{pLL}) and numerics [Fig.~\ref{fig2}(a)]. In the rest of this section, we will show that the leading order discrepancy scales as $n^{3/2}$, where $n$ is the pLL index.

\begin{figure}[tb] 
\includegraphics[width = 8.6cm]{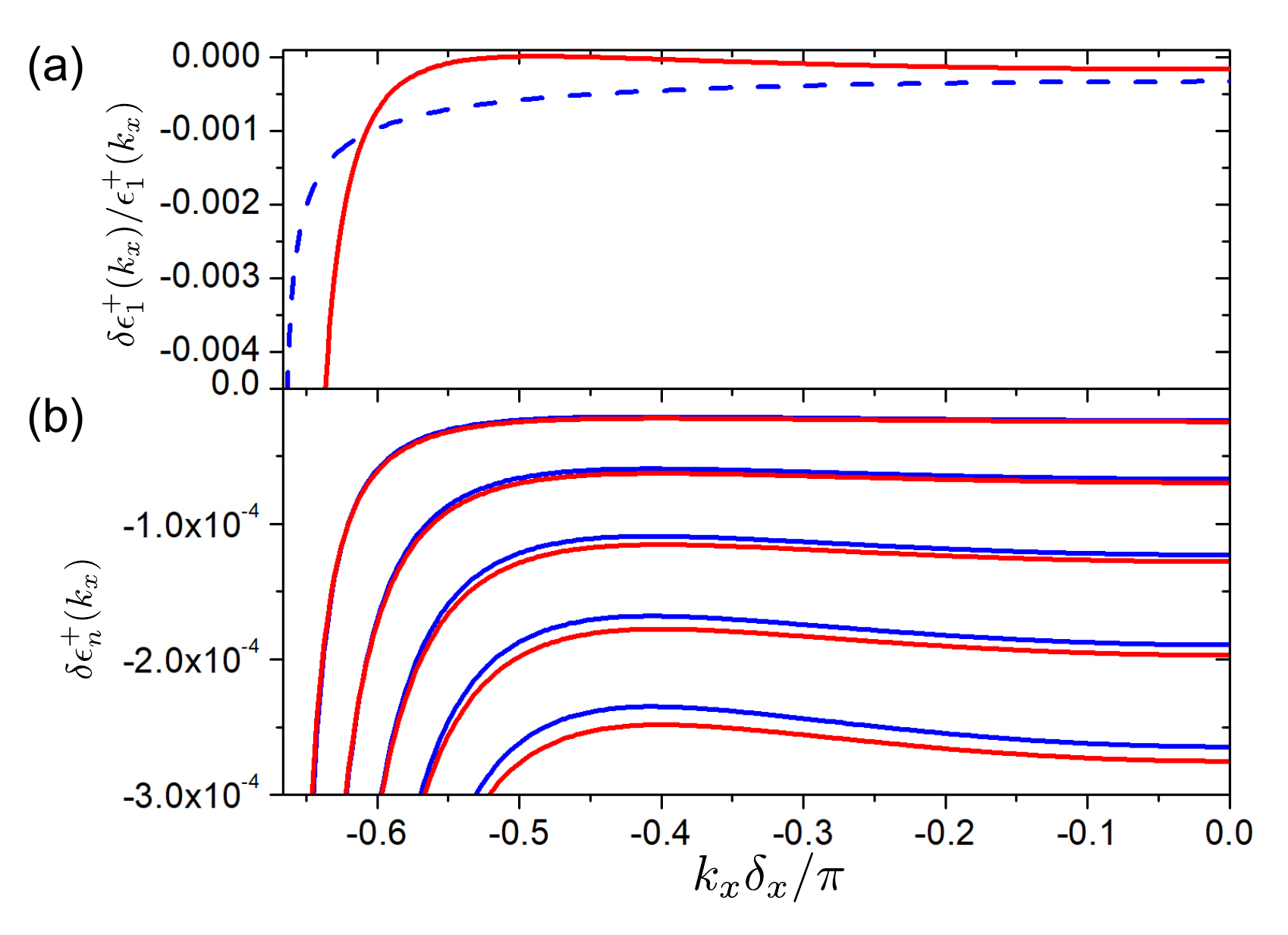}
\caption{Higher-order corrections to the pLL dispersions. As in the main text, $N=1200$, and $\lambda=0.0005a^{-1}$ such that the maximal C-C bond elongation at the edges of the GNR is 27\%. (a) Two leading ($n^{3/2}$) contributions to the relative correction to pLL$_1$, $\delta \epsilon_1^{+}/\epsilon_1^{+}$, as predicted by the continuum theory. The solid red line is the quadratic order contribution $\delta \epsilon_1^{+,(2)}/\epsilon_1^{+}$ [Eq.~(\ref{delta_pLL2})] and the dashed blue line is the cubic order contribution $\delta \epsilon_1^{+,(3)}/\epsilon_1^{+}$ [Eq.~(\ref{delta_pLL3})]. (b) The difference between the numerical pLL dispersions and the Dirac theory [Eq.~(\ref{pLL})], $\epsilon_n^{+,\text{exact}}-\epsilon_n^{+}$ (blue) and the leading corrections to the Dirac theory $\delta \epsilon_n^{+,(2)}+\delta \epsilon_n^{+,(3)}$ (red), with $1\leq n \leq 5$.} \label{fig6}
\end{figure}

Starting from Eq.~(\ref{H_ribbon}), we now expand both $t(y)$ and $\hat{s}_{\pm\delta_y}$ up to the cubic order and treat the quadratic order term $\delta \mathcal{H}^{\text{D},(2)}_{k_x, y}$ and cubic order term $\delta \mathcal{H}^{\text{D},(3)}_{k_x, y}$ as perturbations to the Dirac Hamiltonian [Eq.~(\ref{H_Dirac})]. Explicitly, these terms are
\begin{subequations} \label{delta_H_Dirac}
\begin{align}
\delta \mathcal{H}^{\text{D},(2)}_{k_x, y} &= -\frac{t''(y_0) (y-y_0)^2}{2t(y_0)} t\tau^x + \frac{1}{2}t\delta_y^2 \frac{d^2}{dy^2}  \tau^x, \label{delta_H_Dirac_2}
\\
\delta \mathcal{H}^{\text{D},(3)}_{k_x, y} &= -\frac{ t'''(y_0)  (y-y_0)^3 }{6t(y_0)} t\tau^x - \frac{1}{6}i t \delta_y^3 \frac{d^3}{dy^3} \tau^y, \label{delta_H_Dirac_3}
\end{align}
\end{subequations}
where the $k_x$ dependence is acquired from $y_0$ through Eq.~(\ref{y0}). The matrix elements of such perturbations are easily evaluated using ladder operators, and the perturbation theory calculations are lengthy but straightforward. Specifically, the quadratic order terms in Eq.~(\ref{delta_H_Dirac_2}) contribute a correction
\begin{equation} \label{delta_pLL2}
\delta \epsilon_n^{\pm,(2)}(k_x)=\pm n^{\frac{3}{2}}t \frac{\{3 t''(y_0) t(y_0)+ [t'(y_0)]^2 \} \delta_y^{\frac{3}{2}} t''(y_0)}{|2t'(y_0)|^{\frac{5}{2}} [t(y_0)]^{\frac{1}{2}}},
\end{equation}
which arises from the second-order perturbation theory calculation of Eq.~(\ref{delta_H_Dirac_2}), but the cubic order terms in Eq.~(\ref{delta_H_Dirac_3}) gives rise to 
\begin{equation} \label{delta_pLL3}
\delta \epsilon_n^{\pm,(3)}(k_x)=\mp \frac{3}{16} n^{\frac{3}{2}} t 
\frac{\{ t'''(y_0) \left[ t(y_0) \right]^2 + |t'(y_0)|^3 \} \delta_y^{\frac{3}{2}}}{|t'(y_0)|^{\frac{3}{2}} [t(y_0)]^{\frac{3}{2}}},
\end{equation}
at the first-order perturbation level. Therefore, one cannot naively expect that the quadratic order terms [Eq.~(\ref{delta_H_Dirac_2})] always have a more profound influence than the cubic order terms [Eq.~(\ref{delta_H_Dirac_3})]. In fact, the relative importance between Eqs.~(\ref{delta_pLL2}) and~(\ref{delta_pLL3}) is sensitive to the value of $k_x$ as well as the type of the lattice deformation, though they both scale as $n^{3/2}$. For the twist deformation characterized by Eq.~(\ref{t_sub}), the sizes of these contributions at $n=1$ (relative to the pLL$_1$ dispersion $\epsilon_1^{+}$) are plotted in Fig.~\ref{fig6}(a). Both diverge at the Dirac points, but $\delta \epsilon_n^{\pm,(2)}$ diverges more quickly; in contrast, $\delta \epsilon_n^{\pm,(3)}$ dominates around $|k_x \delta_x|=\pi/2$, and the two corrections become comparable near the $\Gamma$ point. Figure~\ref{fig6}(b) compares $\delta \epsilon_n^{+,(2)}+\delta \epsilon_n^{+,(3)}$ with $\epsilon_n^{+,\text{exact}}-\epsilon_n^{+}$, the difference between the numerical pLL dispersions and the Dirac theory prediction [Eq.~(\ref{pLL})]. The discrepancy can be traced to subleading corrections arising from even higher-order terms in the expansions of $t(y)$ and $\hat{s}_{\delta_y}$. Since the linear order expansions of $\hat s_{\pm \delta_y}$ and $t(y)$ already capture the numerical energy bands to a high accuracy of $10^{-4}t$, which can be further increased by including the quadratic and the cubic order terms, we choose not to include more terms in the expansions of $\hat s_{\pm \delta_y}$ and $t(y)$.

\section{Next nearest neighbor effects}
\label{a4}
Our discussion of the twisted GNR in the main text is based on a tight-binding model with only nearest neighbor hoppings, while the next nearest neighbor effect is in general not negligible in realistic graphene. In this section, we investigate the response of pLLs to the next nearest neighbor hoppings.

With the next nearest neighbor hopping terms included, the strain-free Hamiltonian [Eq.~(\ref{H_inf})] has an additional term
\begin{equation} \label{delta_H_inf}
\delta H_0=-\sum_{\bm r} (t'_i a_{\bm r}^\dagger a_{\bm r + \bm \beta_i}+ t'_i b_{\bm r}^\dagger b_{\bm r + \bm \beta_i}) +\text{H.c.},
\end{equation}
where $(\bm \beta_1, \bm \beta_2, \bm \beta_3) = (\bm \alpha_1-\bm \alpha_3, \bm \alpha_2-\bm \alpha_3, \bm \alpha_1-\bm \alpha_2)$ are three of the six next nearest neighbor vectors. In the presence of an elastic strain, as in Eq.~(\ref{t_sub}), the hopping integrals $t'_i$ \footnote{Note this notation differs from the derivative of the nearest neighbor hopping $t'(y_0)$ by a subscript $i$.} are modulated through
\begin{equation} \label{tp_sub}
t'_i \rightarrow t'_i \exp \{g [1-\tilde \beta_i(\bm r)/\beta_i] \},
\end{equation}
where $\tilde \beta_i(\bm r)$ is the strain-modulated spacing between a chosen lattice site at $\bm r$ and its $i$th next-nearest neighbor and $\beta_i$ is the strain-free counterpart of $\tilde \beta_i(\bm r)$ as illustrated in Fig.~\ref{fig1}(a).

Focusing on the twist deformation, the tight-binding Hamiltonian for the twisted GNR [Eq.~(\ref{H_ribbon})] should be supplemented by
\begin{equation} \label{delta_H_ribbon}
\begin{split}
\delta H= -\sum_{k_x, y}  \psi_{k_x,y}^\dagger \big[ &2 t'_1(y)  (\hat s_{-\delta_y}+\hat s_{\delta_y}) \cos (\tfrac{1}{2}k_x\delta_x )
\\ 
 + &2 t'_3(y) \cos(k_x\delta_x ) \big] \psi_{k_x,y},
\end{split}
\end{equation}
where the hopping integrals along $\bm \beta_{1,2}$ and $\bm \beta_3$ are respectively $t'_1(y) = t' \exp \{ g [1-(1+\tfrac{1}{4}\lambda^2 y^2)^{1/2}] \}$ and $t'_3(y) = t' \exp \{ g [1-(1+\lambda^2 y^2)^{1/2}] \}$ with $t'$ being the isotropic next nearest neighbor hopping in the absence of strain. 

In the lowest order approximation, we may write $\hat s_{-\delta_y}+\hat s_{\delta_y}\approx 2$ so that only onsite terms appear in Eq.~(\ref{delta_H_ribbon}). This observation suggests that the effect of $\delta H$ may be greatly suppressed if we apply an appropriate electric field $E = -\partial_y \phi(y)$, which also induces an onsite potential in the GNR Hamiltonian. Since the pLLs sit at the domain wall $y_0$, we require the external electric potential energy to cancel the onsite energy resulting from the next nearest neighbor hoppings. Therefore, the electric potential at $y_0$ is
\begin{equation} \label{phi_y0}
\phi(y_0)=-\frac{1}{e}[4t_1'(y_0)\cos(\tfrac{1}{2}k_x\delta_x) + 2t_3'(y_0)\cos(k_x\delta_x)].
\end{equation}
Making use of Eq.~(\ref{y0}), we can remove the $k_x$ dependence in Eq.~(\ref{phi_y0}) so that the applied electric potential reads
\begin{equation} \label{phi}
\phi(y) = \frac{1}{e} \bigg[ \frac{2t_1'(y)t}{t(y)} - \frac{t_3'(y)t^2}{t^2(y)} +2t_3'(y) \bigg].
\end{equation}
With this electric field, both the electronic structure and the transport of the twisted GNR are governed by the nearest neighbor effect, which is the concern of the main text. Although the next nearest neighbor effect can be suppressed by an external electric field, the realization of such an electric field is unfortunately not easy due to the complicated space dependence associated with the electric potential $\phi(y)$ in Eq.~(\ref{phi}). In the following, we show that our prediction on the negative strain resistivity and enhanced thermopower should still be qualitatively correct even without the applied external electric field.

By expanding the shift operators $\hat s_{-\delta_y} + \hat s_{\delta_y} \approx 2 + \delta_y^2 \tfrac{d^2}{dy^2}$ and next nearest neighbor hopping $t_{1,3}'(y)$ at $y=y_0$ to the quadratic order of $y-y_0$, we can calculate for the pLL dispersions [Eq.~(\ref{pLL})] the perturbative correction as
\begin{eqnarray} \label{pLLnnn}
&&\delta\epsilon_n^\pm(k_x)=-4 t'_1(y_0)  \cos (\tfrac{1}{2}k_x\delta_x )- 2 t'_3(y_0) \cos(k_x\delta_x ) \nonumber\\
&&+\Bigg\{ -2 \left[ \left. \frac{d^2 t'_1}{dy^2}\right| _{y_0} \frac{t\delta_y}{|\Omega_{y_0}|} - t'_1(y_0) \frac{|\Omega_{y_0}|\delta_y}{t}  \right] \cos (\tfrac{1}{2}k_x\delta_x ) \nonumber\\
&&-  \left. \frac{d^2 t'_3}{dy^2}\right| _{y_0} \frac{t\delta_y}{|\Omega_{y_0}|} \cos(k_x\delta_x ) \Bigg\} \left( n+\frac{\delta_{n0}}{2} \right), 
\end{eqnarray}
which is strongly $k_x$-dependent but only weakly depends on the pLL index $n$. Although Eq.~(\ref{pLLnnn}) is valid for $\tfrac{4\pi}{3\delta_x}\leq k_x \leq \tfrac{8\pi}{3\delta_x}$, one can again shift the domain of $k_x$ into the first BZ so that the $\cos (\tfrac{1}{2}k_x\delta_x )$ terms change sign. The previously flat pLL$_0$ now becomes dispersive, and the particle-hole symmetry is manifestly broken.

\begin{figure}[t]
\includegraphics[width = 8.6cm]{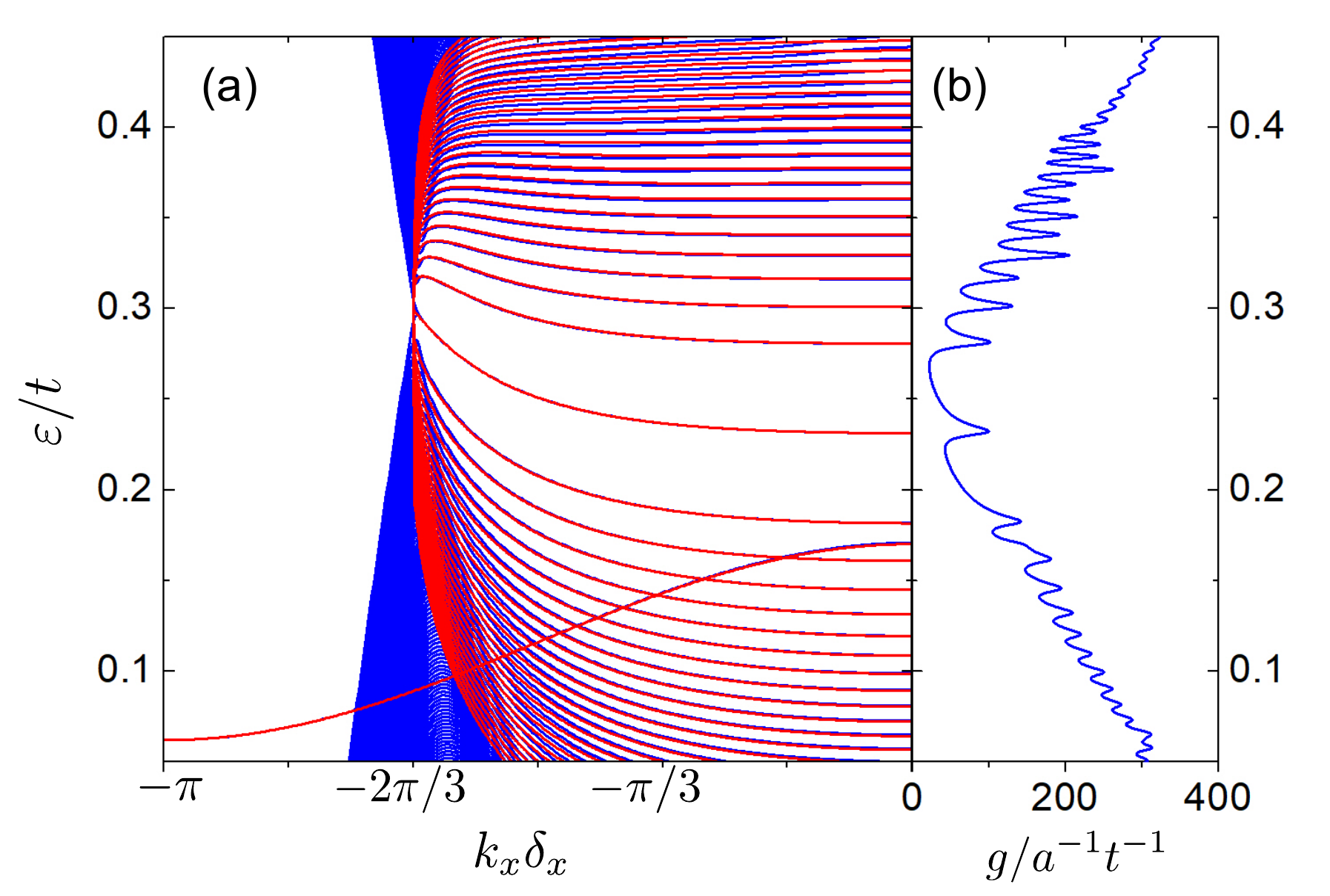}
\caption{Band structure and DOS of a twisted GNR with next nearest neighbor hopping. As in the main text, $N=1200$, and $\lambda=0.0005a^{-1}$ such that the maximal C-C bond elongation at the edges of the GNR is 27\%. The strain-free next-neatest-neighbor hopping is $t'=0.1t$. (a) Low-energy bands from numerical simulations of the tight-binding model (blue) with the analytical predictions (red) of Eqs.~(\ref{pLL}) and (\ref{pLLnnn}) for the pLLs and Eq.~(\ref{esdisp}) for the edge states overlaid. The particle-hole symmetry is broken by the next nearest neighbor hopping, and both the pLL$_0$ and the edge states are now dispersive. (b) The corresponding numerical DOS, broadened by convolution with a Lorentzian of width $\delta_\varepsilon=0.0024t$.} \label{fig7}
\end{figure}

A comparison between our theory for the pLLs and numerical spectrum is given in Fig.~\ref{fig7}(a). We find Eqs.~(\ref{pLL}) and (\ref{pLLnnn}) yield reasonably good agreement for the low-lying pLLs comparing to the existing DFT calculations in Ref.~\cite{zhang2014}. Even better agreement could be achieved by keeping higher-order gradients in Eq.~(\ref{delta_H_ribbon}), as we have done in Appendix~\ref{a3}. As illustrated in Fig.~\ref{fig7}(a), the next nearest neighbor hoppings do not destroy the dispersive pLLs. We thus expect the negative strain resistivity, which results from the intervalley transport mediated by partially filled dispersive pLLs, should persist. We also plot the numerical DOS in Fig.~\ref{fig7}(b); as in the $t'=0$ case, because of the flatness of the pLLs near the $\Gamma$ point, the DOS is sharply peaked and expected to produce an enhanced thermopower there. 

Before we leave this section, we mention that the enhanced thermopower may also arise from the edge states, which are now dispersive and may produce features in the DOS similar to those resulting from the pLLs. It is therefore useful to find the edge state dispersion explicitly. In the framework of the perturbation theory, we first consider the zero mode localized on the $A$ sublattice at the lower edge [see Fig.~\ref{fig1}(a)] by writing the Schr\"odinger equation
\begin{equation}
2t(y_j)\cos(\tfrac{1}{2}k_x\delta_x) \phi_j + t \phi_{j+1} = 0,
\end{equation}
where $y_j=-\tfrac{W}{2}+(j-1)\delta_y$ marks the position of the $a_j$ site and $\phi_j$ is the associated probability amplitude. Note the wave function associated with the $B$ sublattice vanishes near the lower edge. Since we have assumed strong deformation at the edge, $\phi_n$ decays rapidly in the bulk for any $k_x$; thus we can approximate the slow-varying function $t(y)$ by $t(-\tfrac{W}{2})$. The normalized wave function then reads
\begin{equation} \label{eswf}
\phi_{n} \approx [1-4\ell^2 \cos^2 (\tfrac{1}{2}k_x \delta_x )]^{1/2} [-2\ell \cos (\tfrac{1}{2}k_x \delta_x)]^{n-1},
\end{equation}
where $\ell \equiv t(-\tfrac{W}{2}) / t <1/2$. Further approximating $t'_1(y)$ and $t'_3(y)$ by their values at $y=-W/2$, we obtain the edge state dispersion as the expectation value of $\delta H$ in Eq.~(\ref{delta_H_ribbon}),
\begin{equation} \label{esdisp}
\langle \delta H \rangle \approx 4\frac{t(-\tfrac{W}{2})}{t} t'_1(-\tfrac{W}{2}) (1+\cos k_x \delta_x)-2 t'_3(-\tfrac{W}{2}) \cos k_x \delta_x.
\end{equation}
This is degenerate with another edge state localized at the upper edge. As shown in Fig.~\ref{fig7}(a), Eq.~(\ref{esdisp}) is a good approximation to the numerical edge state dispersion, which again traverses the entire BZ and is now well separated from the Dirac point. The edge band notably produces a shoulder on the pLL peak at $\epsilon\approx 0.17t$.

\bibliographystyle{apsrev4-1-etal-title_10authors}
\bibliography{graphene_twist}
\end{document}